\documentclass[a4paper,fleqn,usenatbib]{mnras}

\usepackage{newtxtext,newtxmath}

\usepackage[T1]{fontenc}
\usepackage{ae,aecompl}

\usepackage{graphicx}	
\usepackage{amsmath}	
\usepackage{amssymb}	
 
\usepackage{graphicx,epstopdf}
\DeclareGraphicsExtensions{.ps}
\usepackage[flushleft]{threeparttable}
\usepackage[export]{adjustbox}
\usepackage{caption}
\usepackage{booktabs}
\usepackage{listings}
\usepackage{tikz}
\usetikzlibrary{shapes,arrows}
\tikzstyle{block} = [rectangle, draw, fill=blue!20, 
text width=13em, text centered, minimum height=2em,font=\bfseries,line width=1pt]
\tikzstyle{line} = [draw, -latex', line width=3pt]
\tikzstyle{cloud} = [draw, ellipse,fill=red!20, node distance=3cm,
minimum height=4em]

\title[Mock Observations of SC Models with COCOA]{COCOA Code for Creating Mock Observations of Star Cluster Models}

\author[A.Askar et al.]{
Abbas Askar,$^{1}$\thanks{E-mail: askar@camk.edu.pl}
Mirek Giersz,$^{1}$
Wojciech Pych$^{1}$ and
Emanuele Dalessandro$^{2,3}$
\\
$^{1}$Nicolaus Copernicus Astronomical Centre, Polish Academy of Sciences, 
ul. Bartycka 18, 00-716 Warsaw, Poland\\
$^{2}$INAF - Osservatorio Astronomico di Bologna, Via Gobetti 93/3 40129 Bologna, Italy\\
$^{3}$Dipartimento di Fisica \& Astronomia, Universit\'{\`a} degli Studi di Bologna, Via Gobetti 93/2 40129 Bologna, Italy
}

\date{Accepted XXX. Received YYY; in original form ZZZ}

\pubyear{2015}

\begin{document}
\label{firstpage}
\pagerange{\pageref{firstpage}--\pageref{lastpage}}
\maketitle

\begin{abstract}
We introduce and present results from the \textsc{cocoa} (Cluster simulatiOn Comparison with ObservAtions) code that has been developed to create idealized mock photometric observations using results from numerical simulations of star cluster evolution. \textsc{cocoa} is able to present the output of realistic numerical simulations of star clusters carried out using Monte Carlo or \textit{N}-body codes in a way that is useful for direct comparison with photometric observations. In this paper, we describe the \textsc{cocoa} code and demonstrate its different applications by utilizing globular cluster (GC) models simulated with the \textsc{mocca} (MOnte Carlo Cluster simulAtor) code. \textsc{cocoa} is used to synthetically observe these different GC models with optical telescopes, perform PSF photometry and subsequently produce observed colour magnitude diagrams. We also use \textsc{cocoa} to compare the results from synthetic observations of a cluster model that has the same age and metallicity as the Galactic GC NGC 2808 with observations of the same cluster carried out with a 2.2 meter optical telescope. We find that \textsc{cocoa} can effectively simulate realistic observations and recover photometric data. \textsc{cocoa} has numerous scientific applications that maybe be helpful for both theoreticians and observers that work on star clusters. Plans for further improving and developing the code are also discussed in this paper.

\end{abstract}

\begin{keywords}
globular clusters: general -- galaxies: star clusters: general -- techniques: photometric -- methods: numerical -- methods: observational
\end{keywords}



\section{Introduction}

Due to advancements in computational technology and programming over the past couple of decades, there has been extensive work done in modelling the dynamical evolution of star clusters using sophisticated numerical simulation codes like direct \textit{N}-body and Monte Carlo codes. The long-standing goal of simulating the evolution of a realistic million-body globular cluster (GC) up to twelve-billion years with a direct \textit{N}-body code was recently achieved by \citet{wang16} with the \textsc{nbody6++gpu} code \citep{wang15}. With increasingly more realistic simulations of star clusters, there is also a need to validate and compare results with observations. For this purpose, it is important to present simulation results in a way that would be most useful for direct comparison with observations of star clusters. In this paper, we introduce and present the \textsc{cocoa} (Cluster simulatiOn Comparison with ObservAtions) code\footnote{\url{https://github.com/abs2k12/COCOA}} that can automatically simulate mock photometric observations of star clusters from simulation snapshots and then subsequently apply observational methods and techniques to obtain cluster parameters from the mock images. \textsc{cocoa} will allow theoreticians working with numerical simulations to output their simulated cluster model data as a realistic observation enabling them to get feedback from observers and vice versa. \textsc{cocoa} can create synthetic ideal observations from virtually any optical telescope and imaging camera and has an inbuilt PSF photometry pipeline that can perform photometry on the simulated images and obtain star catalogues by utilizing procedures that are typically employed by observers. \textsc{cocoa} has already been used to create mock observations of \textit{N}-body and Monte Carlo models after 12 Gyrs of cluster evolution in \citet{askar16}, \citet{wang16}, \citet{askar-dark} and \cite{belloni17}.

Numerical simulation codes provide detailed output showing the evolution of global parameters of cluster models and also snapshots containing relevant information for all objects in the star cluster at a specific time. Such detailed output can be extended so that results from simulation models of star clusters can be presented in a format that can be utilized and understood by observers studying objects like GCs. This will not only facilitate a direct comparison between simulated models and observed clusters but will also be effective in determining the accuracy of observational techniques and methods to determine different cluster parameters. The importance of this endeavor to bridge the gap between theoreticians and observers working on star clusters was recognized by \citet{hut02} and efforts were made to simulate synthetic observations of star cluster simulations within the \textsc{starlab} environment \citep{starlab10}. Stellar and binary evolution routines \citep{sse2000,bse2002} implemented within \textsc{NBODY6} and other cluster evolution codes allowed the possibility to obtain magnitudes in few optical filters that could be used to create colour magnitude diagrams (CMDs) from simulation snapshots. Various studies using numerical simulation codes to investigate dynamical evolution and stellar populations in star clusters have employed different methods to compare their results with observations. For instance, \citet{lamers06} used the \textsc{GALEV} code with \textit{N}-body results to investigate the photometric evolution of dissolving clusters, \citet{sippel12} used projected \textit{N}-body snapshots and created FITS images of model GCs to measure their half-light radii, \citet{zhuang15} computed integrated colours and Lick indices for \textit{N}-body models of GCs. Also more recently, \citet{bianchini15} have developed the \textsc{sisco} code to simulate integral field unit (IFU) observations of GCs using data from numerical simulations which opens up the possibility to obtain observationally resolved kinematic measurements of stars in simulated GC models.

An important step towards comparing observations with simulations was the development of the \textsc{MASSCLEAN} code \citep{massclean1,massclean2,massclean3} that could simulate the evolution of star clusters with different initial conditions and create synthetic observational images for those clusters. \textsc{MASSCLEAN} uses stellar evolution tracks to evolve the stars in the cluster model and for dynamics gives the user the option to enable simple linear mass segregation. Although the \textsc{MASSCLEAN} code does not simulate detailed dynamical evolution of the cluster and does not contain binary systems, it can output cluster data in the form of observational images and synthetic CMDs that can be compared with observations. By developing \textsc{COCOA} we seek to continue this work and allow users the possibility to use the results of realistic dynamical simulations from \textit{N}-body and Monte Carlo codes to create observational photometric data that can be automatically reduced with the PSF photometry pipeline provided within the code.

With \textsc{cocoa} it is possible to check if there are any systematics or biases associated with actual observational data and techniques used to determine cluster properties. Our idealized mock observations do not include foreground and background stars, the noise is limited to a Poisson noise, there are no effects of cosmic rays on our images and the PSF does not vary. Actual observations are plagued by all the aforementioned problems. \citet{guidi15} have compared the direct results of galaxy simulations with the observationally-derived quantities from synthetic observations and found that there are systematic differences in obtained galaxy parameters which are in part caused by observational biases. \citet{sollima15} have also shown that fitting analytic models to available observables for star clusters can result in significant biases while determining global parameters such as mass of the cluster.  With \textsc{cocoa} it is possible to investigate such biases and systematics in photometric studies of star clusters. For instance, the influence of photometric errors on determination of cluster parameters like binary fractions can be checked. Accuracy of techniques used to obtain the center of observed GCs can also be tested with \textsc{cocoa}. The completeness of observations can be checked by comparing the results from star catalogues obtained after doing photometry on the simulated images of clusters with the snapshot data that includes information for all stars in the cluster. Moreover, simulating such ideal observations from simulation models will not only provide a critical check on the data reduction processes and techniques used by observers but it will also be helpful in refining initial conditions for simulated cluster models aiming to reproduce Galactic GCs.

Furthermore, \textsc{cocoa} can also be used by observers for the purposes of writing observational proposals based on mock observations of results from realistic numerical simulation models of star clusters. Uncertainties as to how simulated cluster models would appear if they were observed as a real clusters on the sky can also be checked. Observational analysis of simulated cluster models may reveal numerical issues and problems with erroneous initial conditions of simulated models. As mentioned earlier, a thorough comparison between simulated models and observations can help better constrain initial model, parameters that control stellar evolution, binary parameter distribution, mass functions and other initial structural assumptions that are used by theorists. This makes \textsc{cocoa} an important tool that may help observers in assisting theoreticians to improve their theoretical models.

In this paper we will mainly concentrate on describing \textsc{cocoa} and using its basic features for creating mock observations of results from \textsc{mocca} code \citep[and reference therein]{Gierszetal2013} for star cluster simulations and generating observational CMDs after carrying out PSF photometry on mock images. More sophisticated \textsc{cocoa} applications for generating surface brightness profiles, velocity dispersion profiles, number density profiles, luminosity functions and fitting them to different models will be discussed in future papers. \textsc{cocoa} has been primarily developed in Python and it combines various tools in other programming languages to offer different functionalities. In the second section of this paper, we provide a description of the \textsc{cocoa} code and how it works. In the third section we show results from \textsc{cocoa} by creating mock observations of cluster models that were simulated using the \textsc{MOCCA} code. 

In the same section, we also present comparisons with an observation of the Galactic GC NGC 2808 and other different applications of the \textsc{cocoa} code. In the fourth section we discuss future developments and modules that will be developed for applying various techniques to the synthetic observational data. An appendix with a brief manual for the code is also provided.

\section{Creating Mock Observations of Star Clusters}

\begin{figure*}
\begin{center}
  \begin{tikzpicture}[node distance = 2.5cm, auto]
    \node [block,text width=15em] (initsnap) {\Large Simulation Snapshot};
    \node [block, below of = initsnap,text width=15em] (projection) {\Large Projection Module  \\ (Section \ref{sec:projection})};
    \node [block, below of = projection,text width=15em] (proj-snap) {\Large Projected Snapshots with Magnitudes (Section \ref{sec:galevnb})};
    \node [block, below of = proj-snap, node distance = 3cm,text width=15em] (FITS) {\Large Synthetic FITS Image of the Observation};
    \node [block, below of = FITS,node distance = 3cm,text width=25em ] (final-cat) {\Large Catalogue of All Observed Stars \\ (Positions, Magnitudes, Error)};
    \node [block, left of = projection, node distance = 7cm,text width=15em] (input-file) {\Large Input File for\\ Projection Module \\(Appendix \ref{sec:app-projection})};
    \node [block, left of = proj-snap, node distance = 7cm,text width=15em] (observation) {\Large Imaging/Observation Module \\ (Section \ref{sec:sos1})};
    \node [block, left of = FITS, node distance = 7cm,text width=16em] (sim2obs) {\Large Input File for Imaging Module\\ - \textsc{sim2obs} parameters \\ (Appendix \ref{sec:app-params})};
    \node [block, right of = FITS, node distance = 5.5cm, minimum height=2em,text width=14em] (photometry) {\Large Photometry Module \\ (Section \ref{sec:photo} and \ref{sec:phot-pars-dao})};
    \path [line] (initsnap) -- (projection);
    \path [line] (projection) -- (proj-snap);
    \path [line] (proj-snap) -- (FITS);
    \path [line] (FITS) -- (final-cat);
    \path [line] (input-file) -- (projection);
    \path [line] (proj-snap) -- (observation);
    \path [line] (sim2obs) -- (observation);
    \path [line] (photometry) -- (FITS);
  \end{tikzpicture}
 \caption{\textsc{cocoa} Flow Chart: A schematic diagram of the main steps and the sequence of the code.}
   \label{flow-chart}
\end{center}
\end{figure*}
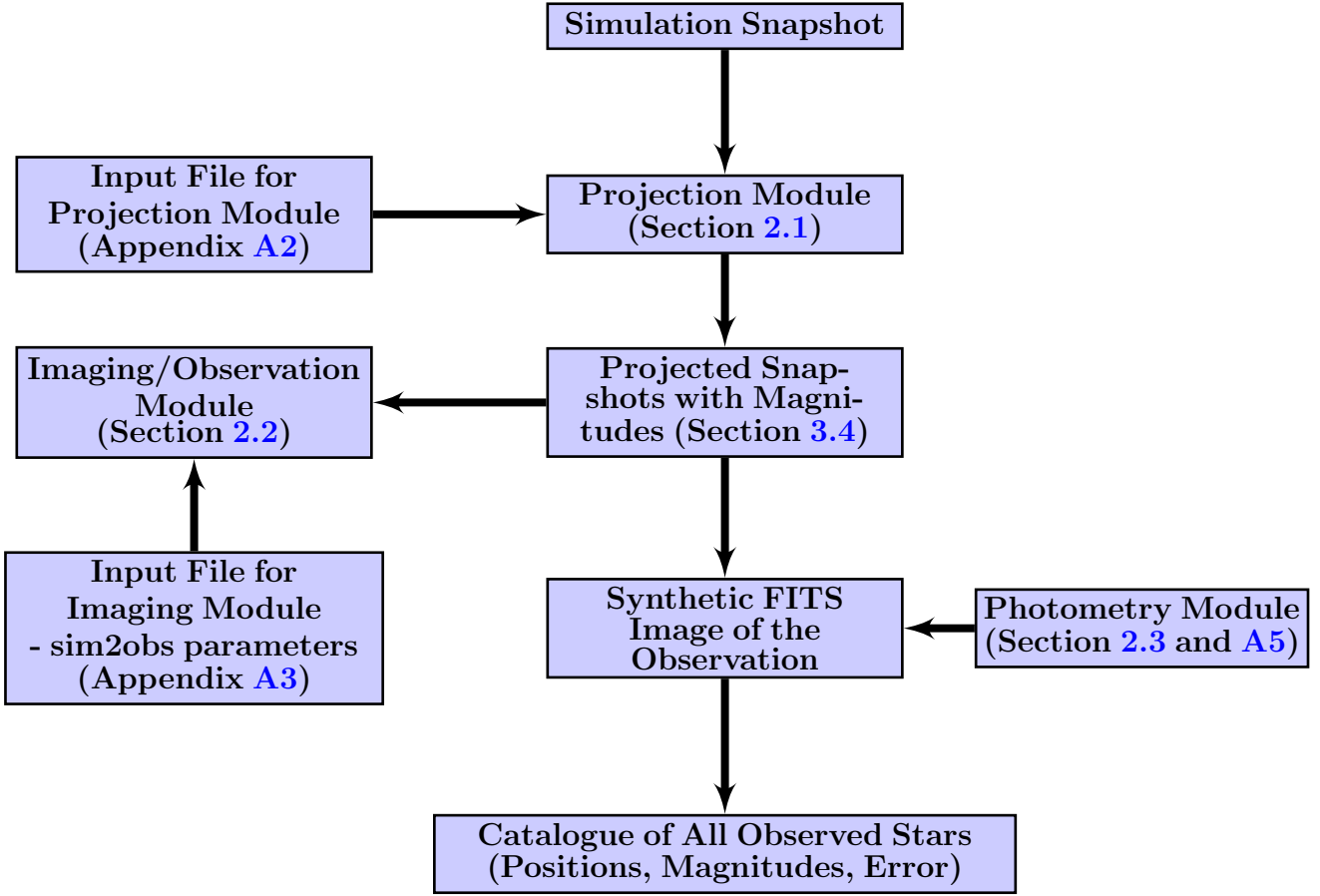

\textsc{cocoa} was designed in a way that would be most user-friendly and with as many automated procedures as possible to assist users that are not familiar with observational reduction procedures. \textsc{COCOCA} comes with default values and a working example that shows how to use the different features of the code. The \textsc{cocoa} code has essentially three modules. The first one is the projection module that projects numerical snapshot data for all stars in the cluster model on to the plane of the sky. The procedure and features of the projection module are provided in section \ref{sec:projection}. The second module simulates the observation using the projected snapshot. The user provides the module with the parameters of the instrument/telescope with which they want to simulate the observations and it automatically generates the required number of frames to create a composite image of the entire cluster. The details of this module are provided in sections \ref{sec:sos1}. The third module is the automated PSF photometry pipeline in \textsc{cocoa} that carries out photometry on the simulated images produced using second module andthe output from the PSF photometry code is used to create a catalogue of all observed stars in the cluster. The details of this module are provided in section  \ref{sec:photo}. A flow diagram illustrating the main steps of the code is shown in Figure \ref{flow-chart}.

The simulation results used to generate mock observations in this paper come from the \textsc{MOCCA} code for star cluster simulations. \textsc{MOCCA} is based on the orbit averaged Monte Carlo method of \citet{henon71} for following the long term evolution of spherically symmetric star clusters which was further improved by \citet{stdk86} and \citet{Gierszetal2013}. Monte Carlo codes for star cluster evolution are more approximate than direct \textit{N}-body codes but they are significantly faster. \textsc{MOCCA} uses the \textsc{fewbody} code \citep{freg04} to compute strong binary-single and binary binary interactions. For stellar and binary evolution, \textsc{MOCCA} uses SSE and BSE \citep{sse2000,bse2002}. \textsc{MOCCA} was recently used to carry out a survey of about two thousand star cluster models, \textsc{MOCCA}-Survey Database I \citep{askar16} which samples a wide array of initial parameters that are important in determining the evolution of cluster models. Results from \textsc{MOCCA} have been extensively compared with \textit{N}-body simulations and the agreement for both the evolution of global parameters and number of specific objects and binary systems is remarkable \citep{Gierszetal2013,wang16,madrid17}. For using \textsc{COCOA}, one needs two dimensional positions and absolute magnitudes of stars that could be provided by either Monte Carlo or \textit{N}-body codes.

\subsection{Projecting Simulation Snapshot}\label{sec:projection}

As an input, \textsc{cocoa} requires a snapshot at any given time from numerical star cluster simulations. The results presented in this paper utilize the snapshots from the \textsc{mocca} code. These snapshots from \textsc{mocca} simulations are produced periodically during the cluster evolution and contain the details of all the objects in the GC model at a given time, these include their spatial,kinematic and stellar evolution properties (provided by the \textsc{sse} \citep{sse2000} and \textsc{bse} \citep{bse2002} codes used by \textsc{mocca}) like position from the center of the cluster, velocities, mass, radius, bolometric luminosity, magnitudes in three filters (B,V \& I) and other important parameters like semi-major axis and eccentricity for binaries. \textsc{cocoa} can project the positions and velocities of the star on the plane of the sky and decompose the positions and velocities of individual stars in binary systems. The procedure for decomposing the binaries uses the semi-major axis of the relative orbit,its eccentricity and masses of stars to compute semi-major axes of the barycentric orbits of the individual stars. The projected snapshot which \textsc{cocoa} outputs automatically provides additional parameters like the Keplerian orbital elements for binary systems that are taken from a random distribution functions. Using one simulation snapshot it is also possible to use \textsc{cocoa} to simulate additional projected snapshots on shorter user-defined time scales (hours, days to years) in which the position and velocities of individual stars in binaries can be tracked by \textsc{COCOA}. The projection module can also be used to rotate and change the orientation of the observed cluster model. This can be very valuable in understanding the influence of how viewing a star cluster at a particular orientation influences the derived observational parameters.

\subsection{Imaging Cluster Models} \label{sec:sos1} 

Using 2D positions from our projected snapshot, the \textsc{sim2obs} \footnote{Developed for \textsc{COCOA} by Wojciech Pych in the programming language C. Can also be used as a standalone tool (see Appendix \ref{sec:app-params}).} tool in \textsc{cocoa} is utilized to generate FITS images of simulated cluster models. The \textsc{sim2obs} tool requires an input parameter file and the projected snapshot (from the projected snapshot, \textsc{sim2obs} requires the x,y position of each star and its absolute magnitude) to run. The input parameter file for \textsc{sim2obs} requires a number of parameters that primarily define the properties of the mock observation, particularly the telescope, the imaging camera, the PSF model and distance to the cluster model. The properties for the instrument and the imaging camera include the size of the image, exposure time, saturation level, gain etc. A full list and explanation of these parameters is provided in the appendix \ref{sec:app-params}. Using these user entered parameters \textsc{sim2obs} computes the flux from each star and the corresponding number of counts to accurately produce synthetic FITS images of the star cluster model. Another important parameter which \textsc{sim2obs} requires is image offset values from the center of the cluster. When offset values are zero then the center of the synthetic mock observation is always the cluster center. By changing the offset values, it is possible to point towards a specific part of the cluster in order to observe it.

The \textsc{sim2obs} tool can be used to image an entire star cluster model up to its effective radius by creating multiple mock observations by changing the offset parameter in \textsc{sim2obs}. A procedure has been devised in \textsc{cocoa} to automatically create multiple frames that collectively image the entire cluster model as a mosaic. When projecting the simulation snapshot, the code determines the effective radius of the cluster which is then converted to arcseconds using the user entered distance to cluster. Using the pixel scale value and the image size provided by the user, the code can determine the total number of frames that will be required to image the cluster from the center up to the limiting radius. An overlap factor can also be added by the user to make sure stars at the edge of the frame are accounted for in the neighbouring frame. Offset values are determined and multiple parameter files are written for \textsc{sim2obs} that are then run through a script to create FITS files for different frames which when combined image the entire cluster. The FITS image files can be viewed with astronomical image viewing applications such as SAOimage DS9, CASA viewer, or Starlink Gaia. These applications can also be used to export the FITS image in commonly used picture formats.

\subsection{Automated PSF Photometry}\label{sec:photo}

Once FITS images of different frames that collectively observe the entire cluster star cluster model have been created (please see Appendix \ref{sec:app-params} for more details of this procedure), it is then possible to carry out photometry on these files to recover positions and magnitudes of stars in the same way as observers would do. In order to do this, a PSF photometry pipeline that uses \textsc{daophot/allstar} \citep{stetson87,stetson94} standalone program have been developed in \textsc{cocoa}. The automated photometry pipeline in \textsc{cocoa} makes use of wrappers to call \textsc{daophot/allstar}. This approach is, in part, similar to the photometry pipeline developed by the \textsc{smarts} \textsc{xrb} team \citep{buxton} for carrying out photometry of optical and infrared observations of X-ray binaries.  The following steps are taken to perform the PSF Photometry.

\begin{enumerate}
 \item The photometry module goes through each FITS file of an observed frame one by one. The first step is to determine the FWHM value of the PSF from the mock observation of the image. This value of the FWHM in the simulated images is given by the ratio between the seeing and the pixel scale value that gives the spatial resolution of the mock observation (please see Appendix \ref{sec:app-params}). This value can be directly provided to the module or one may use a stand-alone program \footnote{This program to find the FWHM was developed by Wojciech Pych and has been used in determining FWHM of actual observational images of star clusters for the Cluster AgeS Experiment (CASE) \citep[and references therein]{pych2001,casepaper,rozyczka17}.} that is provided with \textsc{cocoa} to obtain the FWHM of the FITS file.
 \item The value for the FWHM is then used to prepare the option files required by \textsc{daophot} and \textsc{allstar} to carry out PSF photometry. The value of the FWHM and the values entered by the user while generating the synthetic observations are used to determine the parameters required in the options file. These parameters and how they depend on the FWHM is given in the appendix \ref{sec:phot-pars-dao}.
\item After the option files to prepare \textsc{daophot} and \textsc{allstar} are written, the photometry module creates a wrapper for \textsc{daophot} by writing a bash script that will automatically run the code. The wrapper is called with the image name, this allows \textsc{daophot} to know which file to process. Using the standard commands in \textsc{daophot}, we obtain a PSF model and all the required output files from  \textsc{daophot} that are needed to do our PSF photometry. These standard commands involve attaching the image file and then carrying out aperture photometry on the image. The \textit{pick} command in \textsc{daophot} is then used to identify stars to create the PSF model. The number of stars to be picked and their limiting instrumental magnitude is provided in the wrapper. Three iterations of the \textit{psf} command in \textsc{daophot} are used to get a reliable PSF model.
\item Once the PSF model is generated, another wrapper is created to run \textsc{allstar} via a bash script. The required input file names are provided and then the script is executed. The \textsc{allstar} code runs and outputs the final catalogue. The \textsc{cocoa} piepeline reads the contents of the catalogue file produced by \textsc{allstar}  and stores the star number, x coordinate of the position, y coordinate of the position, magnitude and error in the magnitude in an array. The positions in the catalogues produced by \textsc{allstar} are in the units of pixels. The code automatically converts the positions to arcsecond (using the pixel scale value entered by the user when FITS images were being generated) and then using the offset values it corrects the x and y positions in arcseconds so that we have the positions with respect to the center of cluster. The corrected positions, magnitudes and error in magnitude are written to an output file.
 \item The code then moves on to the next FITS image and repeats the steps 1 to 4 and the catalogue from each FITS image is appended to our output file.
 \item Once all \textsc{daophot} and \textsc{allstar} have run on all the frames that image the entire cluster, the code then runs the catalogue through a script that cleans out the overlapping stars. This script compares position of all stars with each other and if there are duplicate stars in the catalogue that came from two different frames then for simplicity, the one with the lower photometric error is selected. After cleaning for overlap stars, we get a final clean catalogue of all stars in the GC.
\end{enumerate}

\begin{table*}
\centering
\resizebox{\textwidth}{!}{%
\begin{tabular}{@{}ccclccc@{}}
\toprule
\textbf{Model} & \textbf{\begin{tabular}[c]{@{}c@{}}Initial Number\\  of Objects\end{tabular}} & \textbf{\begin{tabular}[c]{@{}c@{}}Initial Mass\\  ($M_{\odot}$)\end{tabular}} & \textbf{\begin{tabular}[c]{@{}l@{}}Initial Binary\\  Fraction (\%)\end{tabular}} & \textbf{\begin{tabular}[c]{@{}c@{}}Galactocentric\\ Radius (kpc)\end{tabular}} & \textbf{\begin{tabular}[c]{@{}c@{}}12 Gyr Mass\\ ($M_{\odot}$)\end{tabular}} & \textbf{\begin{tabular}[c]{@{}c@{}}12 Gyr Binary\\ Fraction (\%)\end{tabular}} \\ \midrule
Tidally-Filling & $1.12 \times 10^{6}$ & $8.07 \times 10^{5}$ & \multicolumn{1}{c}{95} & 8.0 & $3.86 \times 10^{5}$ & 28.8 \\ \bottomrule
\end{tabular}%
}
\caption{Initial properties, mass and binary fraction at 12 Gyrs for the model for which the CMD is shown in Figure \ref{cmd1-new}. The initial mass function (IMF) for this model is a three segment Kroupa IMF \citep{kroupa1993}. Black holes and neutron star natal kicks were drawn from  from a Maxwellian distribution with $\sigma$=265 km/s \citep{hobbs2005}.}
\label{tab-model-1}
\end{table*}

\subsubsection{Magnitude Corrections \& Using Single PSF models for Multiple Frames}
It is required to convert instrumental magnitudes to apparent or absolute magnitudes for meaningful comparisons. In order to do this, we prepare a list of around a hundred well separated single stars with a range of different absolute magnitudes. We created a mock observation for these hundred stars by using the exact same parameter files that were used to image the cluster models. The stars are placed at the same distance as the cluster model and have been given artificial positions that form a grid structure on the projected plane. The stars are well separated from each other in order to avoid problems due to crowding/blending. The code carries out photometry on this frame and recovers positions and instrumental magnitudes of stars. The positions of the stars obtained after photometry are converted to parsec (from pixels) and then matched with the positions of the stars in the original list of the stars in order to compare the obtained instrumental magnitude for those stars with their actual absolute magnitudes. Using the differences between the instrumental magnitude and the actual magnitude of the stars, a linear fit can be done which provides us with a correction term to convert our instrumental magnitude to absolute or apparent magnitudes. 
This procedure is also useful in getting a good PSF model for a very dense cluster model. As the stars are well separated in this image, the PSF model obtained during the photometry of this frame has a lower error value and this PSF model can then be used to fit all the frames that collectively image the simulated cluster models. This gets rid of the errors associated with differences in PSF models obtained when numerous frames are required to image the entire cluster and the FWHM and PSF is computed individually for each frame.

\section{COCOA in Action: Applications \& Results}


\subsection{Mock Observations \& Photometry of a Cluster Model}

In this section, we present results from \textsc{cocoa} for a tidally-filling cluster model at the time 12 Gyr. The model was evolved using the \textsc{mocca} code. The snapshot at 12 Gyr was used as an input file in \textsc{cocoa} and the projected snapshot with positions, velocities and magnitudes with respect to the the center of the cluster was created. Then \textsc{cocoa} and the \textsc{sim2obs} tool in it were used to automatically image the entire cluster. Photometry using our automated pipeline was carried out and colour-magnitude diagram were created for the star cluster. In this section, the details of how mock observations are generated and used to obtain photometric results are provided.

From the results of MOCCA survey, a star cluster model with initially $N=1.12 \times 10^6$ objects, low concentration, high initial binary fraction of 95\% and a metallicity of $Z = 0.001$ was selected. High initial binary fraction with properties given by the Kroupa IBP \citep{kroupa95} for GC models can reproduce observed present-day binary fraction as well as the observed anti-correlation between the binary fraction and cluster mass \citep{leigh15,belloni17}. Initial conditions for this model were not explicitly selected to reproduce a particular Galactic GC, the main idea was to look at the evolution of stellar/binary populations in a sparse cluster in which dynamical interactions were not too frequent. The initial binary parameter distribution for this model used the Kroupa IBP for the semi-major axis and eccentricity distributions but had an almost flat distribution for the mass ratio. The model was simulated in an external potential assuming a point-mass potential for the Milky Way at a Galactocentric radius of 8 kpc. At 12 Gyr, the model has $N=9.67 \times 10^5$ objects with a 28.8\% binary fraction and is extended up to 60 pc. The high binary fraction at 12 Gyr is dominated by binaries in which at least one component is less massive than $0.3 M_{\odot}$. Initial and 12 Gyr properties for the model are provided in Table \ref{tab-model-1}. We used the projected x, y positions and the magnitudes for each star in V-band to generate synthetic observations of the cluster. The cluster was observed at a distance of 5 kpc. In the next subsection (\ref{sec:tele-par}, the details of the parameters that were used to simulate the observations for this cluster are provided.

\subsubsection{Telescope Parameters}\label{sec:tele-par}

The parameters of instrument with which the mock observations were created was extremely idealized with very low seeing values, high spatial resolution and a Gaussian PSF. Instrument specifications were chosen to match those of a typical 8-meter class telescope. The pixel scale that we specified in \textsc{sim2obs} was 0.10 arcseconds/pixel and the size of the CCD was 4096 $\times$ 4096 pixels. In order to image the entire cluster up to a radius of 60 pc from the center of the cluster, the \textsc{cocoa} code automatically generated offsets (using the size of the CCD, pixel scale and distance to the cluster) and parameter files for sim2obs. The code then created 169 FITS files in order to effectively image the cluster up to a radius of 60 pc ($\sim 2475$ arcseconds). The spatial resolution of this simulated instrument is fairly high and the FOV of each frame is 409.6 arcseconds. The saturation level for the synthetic observation was kept artificially large in order to avoid too many saturated pixels and have as ideal observations as possible for high exposure times. The important parameter values provided to sim2obs in order to image the cluster in the V-band filter are given below (see Appendix \ref{sec:app-params} for more details about these parameters):
\begin{itemize}
  \item DISTANCE = 5000 [parsecs]
  \item PIXSCALE = 0.10 [arcseconds/pixel]
  \item GAIN = 0.9 [photons/ADU]
  \item SATLEVEL = 990000.0 [counts]
  \item EXPOSURE = 35 [multiple of direct counts]
  \item SEEING = 0.3 [arcseconds]
  \item BACKGROUND = 1
  \item PSF = G [Gaussian profile was used for the PSF Model] 
  \item NOISE = 1 [Poisson noise was enabled]
\end{itemize}

Similar values were used for generating the B-band images. However, a higher exposure value of 65 (multiple of direct counts was used) as stars are dimmer in the B-band and require longer exposure to be imaged. Also when using the \textsc{daophot} routine to pick the stars for making the PSF model, a brighter limiting magnitude for B-band was used. 


\subsubsection{V-band \& B-band Photometry}

Once the FITS images were produced for both the V-band and B-band filters, the automated PSF photometry pipeline in \textsc{cocoa} was used to create a catalogue of all stars that contains their magnitudes and (error in magnitudes) using the procedures explained in Section \ref{sec:photo} and Appendix \ref{sec:phot-mod} and \ref{sec:phot-pars-dao}. The values in the option files for \textsc{daophot} and \textsc{allstar} for these models can be found in Table \ref{option-params}.

\begin{figure}
 \includegraphics[width=\columnwidth]{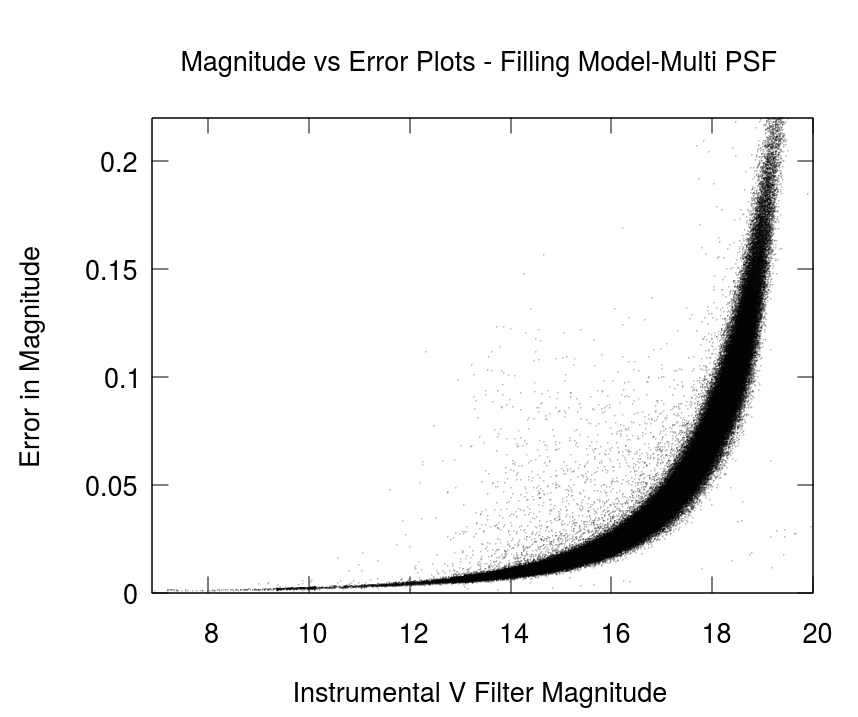}
 \includegraphics[width=\columnwidth]{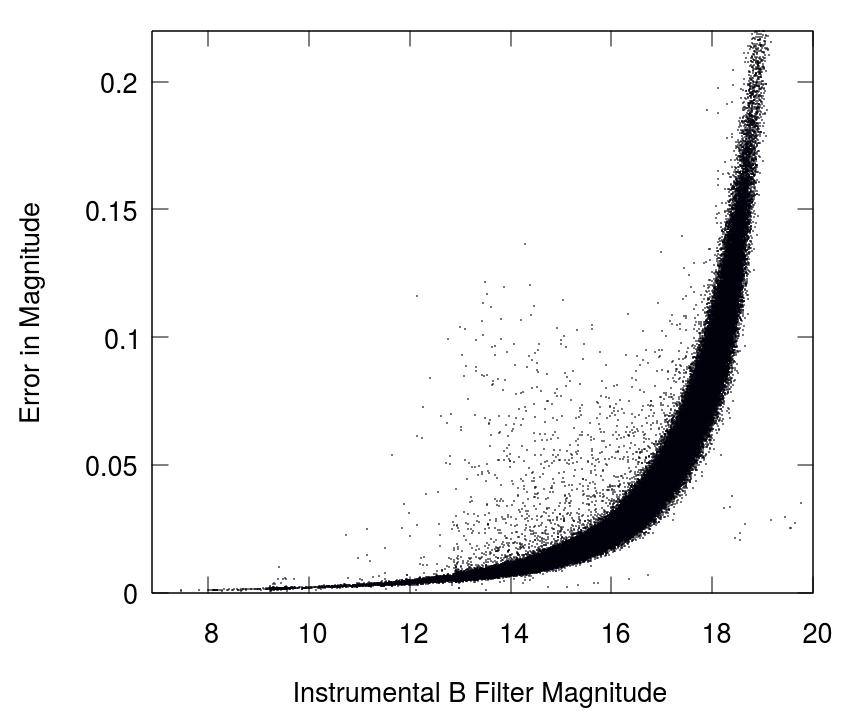}
 \caption{Error vs magnitude plot for V-band and B-band photometry. The x-axis represents the instrumental magnitude and the y-axis represents the error in magnitude.}
 \label{fig:mag-errors}
\end{figure}

From photometry of our simulated FITS images, we were able to recover 201,005 stars in the V-band and 147,966 stars in the B-band. From the simulation snapshot, we know that the actual number of stars were 966,998. The number of recovered stars depends significantly on the exposure value that is used. More stars can be recovered with higher exposure values. Figure \ref{fig:mag-errors} shows the magnitude vs error plots in V and B band filters obtained from the PSF photometry of the mock observations.

\begin{table*}
  \centering
    \begin{tabular}{rrr}
    \toprule
    Parameters in daophot.opt &
      Parameters in photo.opt &
      Parameters in allstar.opt
      \\
    \midrule
    FWHM=3.02 &
      A1=4.53 &
      fit=3.624
      \\
    FIT=3.624 &
      IS=12.08 &
      isky=6.04
      \\
    PSF=8.154 &
      OS=15.1 &
      osky=15.1
      \\
    READ=0.1 &
       &
      watch=0
      \\
    GAIN=0.9 &
       &
      redet=1
      \\
    TH=5.0 &
       &
      
      \\
    AN=1 &
       &
      
      \\
    LOWBAD=15 &
       &
      
      \\
    HIBAD=990000.0 &
       &
      
      \\
    WATCH=-2.0 &
       &
      
      \\
    VAR=0 &
       &
      
      \\
    \bottomrule
    \end{tabular}%
    \caption{Values of parameters in \textsc{daophot/allstar} option files for the test model. The values in these option files were generated using the prescriptions explained in Section \ref{sec:phot-pars-dao}. The FWHM of our simulated images was 3.0.}
  \label{option-params}%
\end{table*}%

\subsubsection{Colour Magnitude Diagram}

\begin{figure}
\begin{center}
 \includegraphics[width=\columnwidth]{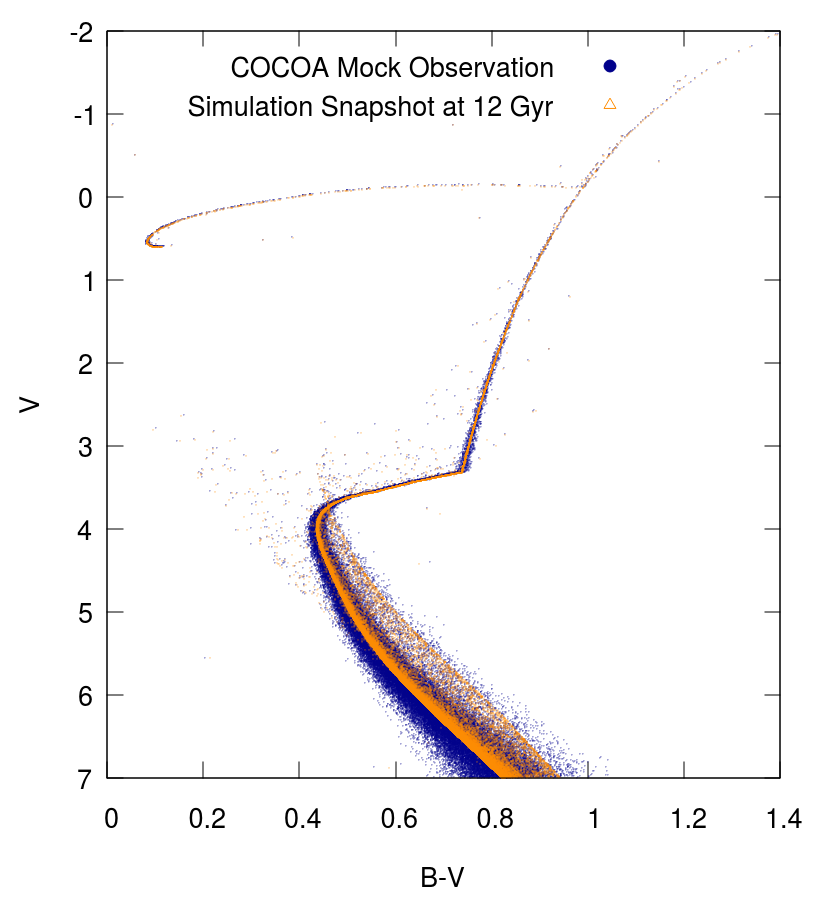} 
 \caption{Colour Magnitude Diagram (CMD) for the cluster model at 12 Gyrs.The magnitudes from observations have been scaled to absolute magnitudes by comparing it with the CMD taken from the simulation snapshot. The B-V colour is on the x-axis and the V-band magnitude is on the y-axis. The blue points indicate observed stars from our PSF photometry pipeline. The orange points are from the CMD directly taken from the snapshot of our simulation results. }
   \label{cmd1-new}
\end{center}
\end{figure}

Once the catalogues from the V-band and B-band photometry were obtained, we joined the data for stars which had the same positions (within 0.01 arcseconds) in both the catalogues. The joined data allowed us to create a colour-magnitude diagram of our model (Figure \ref{cmd1-new}). We compared this observed CMD with the one obtained directly from the simulation data and we can see that the observed CMD well recovers the simulation CMD till up to 3-4 magnitudes below the MS turn-off.


\subsection{Testing Photometry Pipeline with Actual Observations}\label{sec:compare}

In order to ensure that the automated photometry module was working accurately and to check the validity of the parameters used in the \textsc{cocoa} code for working out synthetic images, we used the photometry module to carry out photometry on actual B and V band observations of a real star cluster. Photometry was also performed on the same observations independently by observers using their own parameters.

\begin{figure}
\begin{center}
 \includegraphics[width=\columnwidth]{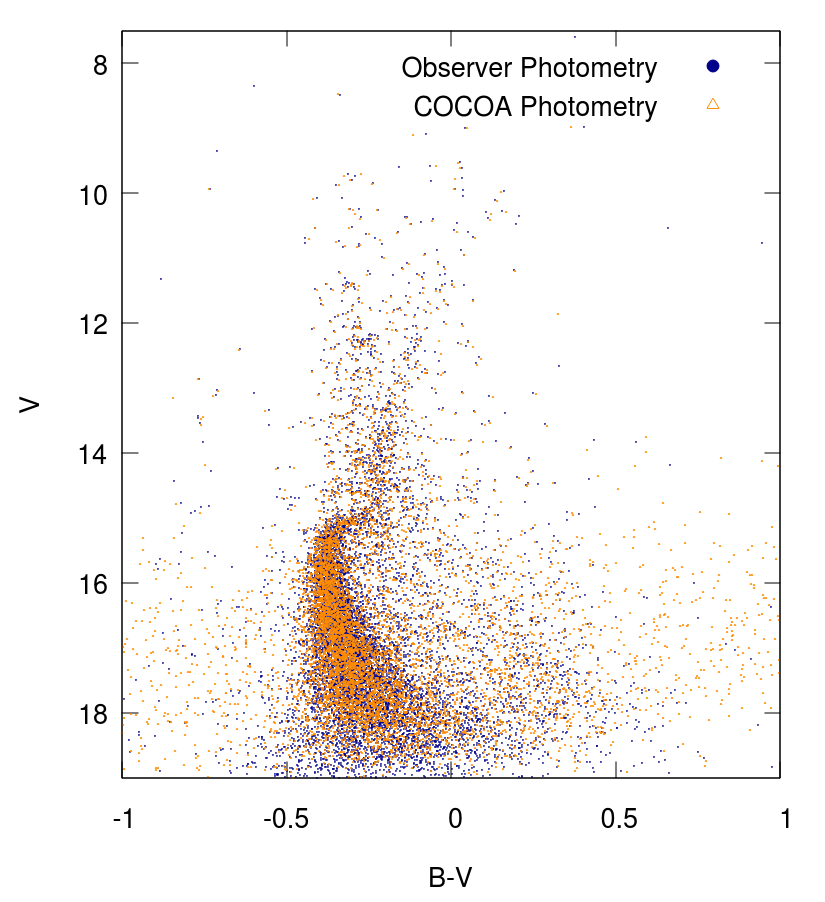} 
 \caption{Colour magnitude diagram for an actual observation of the Galactic GC NGC 2808. The magnitudes are instrumental and the orange points show the results from the module in COCOA while the blue points are the photometry results done by observers.}
   \label{compare-real}
\end{center}
\end{figure}

The observations were made using the ESO Wide Field Imager (WFI) mounted on the 2.2 meter telescope in La Silla, Chile. The particular observations in V and B bands imaged part of the Galactic GC NGC 2808. Photometry was done on both the V and B band images using the automated photometry module in \textsc{cocoa}. A CMD was created and compared with the CMD obtained independently by observers \citep{ema2011}. Figure \ref{compare-real} shows the two CMDs for the same observation, the orange points are from the photometry results from \textsc{cocoa} and the blue points are from the photometry done by the observers. We also compared the photometric errors from the two different procedures and found that \textsc{cocoa} can do reasonably well photometry with its automated pipeline. There is a scatter in the CMD obtained from \textsc{cocoa} results that can be seen with the orange points in Figure \ref{compare-real}. This scatter is connected with the incorrect matching of stars from the catalogues in two different filters. Nearly 20\% of the stars from both catalogues were not properly matched, nearly all these stars were from the dense part of the observed image (left panel in Figure \ref{ngc2808-cocoa}). The match in the CMD can be  made more accurate by improving the criteria to match stars from the photometry results from 2 different filters in \textsc{cocoa}. The matching routines will be further improved to take into account the magnitudes of stars in the next version of the \textsc{cocoa} code. 

\subsection{Mock Observations of a Cluster Model Similar to NGC 2808 }\label{sec:ngc2808}

In the previous subsection, we showed that the photometry pipeline in \textsc{cocoa} can be used to reduce observational data. In this subsection, we compare the observation of NGC 2808 with simulated mock observations of a star cluster model created using \textsc{cocoa} with the same age and metallicity as NGC 2808. We use the snapshot at 11 Gyr from a globular cluster model that was simulated using the \textsc{mocca} code and has total luminosity and tidal radii that are comparable to NGC 2808 in order to simulate mock observations with \textsc{cocoa}. The model was initially selected from a database of 1948 simulated cluster models as part of the MOCCA-Survey Database I \citep{askar-gw} project. The model was then re-simulated with a larger number of stars ($N=2.2 \times 10^6$) and the metallicity for the simulated cluster model was set to [Fe/H]=-1.18 ($Z=0.00132$) to match the observed metallicity of NGC 2808 \citep{carretta09}. For such a comparison, the metallicity value provided to the \textsc{sse} and \textsc{bse} codes for binary stellar evolution in \textsc{mocca} is important, as the fundamental properties of the star like mass, radius and luminosity that will determine their magnitudes depends significantly on this value and the age of the cluster.

The 11 Gyr snapshot was taken from the output of the \textsc{mocca} simulation. The projection module in \textsc{cocoa} was used to project this snapshot. The properties of the stars (mass, luminosity and effective temperature) from the simulation snapshot were provided to the \textsc{GALEVNB} code (explained in the next subsection \ref{sec:galevnb}) to obtain absolute magnitudes in ESO WFI specific V/89 and B/99 filters for each star. Using these magnitudes and the projected positions of the stars, \textsc{cocoa} was used to create mock observations that were simulated with parameters set to replicate the ESO WFI mounted on a 2.2 meter telescope. The FOV was offset to approximately replicate the observed pointing and the distance was set to 9.6 kpc to match the observed distance to NGC 2808 \citet[][updated 2010]{Harris1996}. Observations were simulated in B and V band with the same exposure times and seeing value as the actual observations. Figure \ref{ngc2808-cocoa} shows the actual observational image (for the V filter) that was used for this comparison and the mock image of the cluster model at 11 Gyrs created with MOCCA. The parameters provided to the sim2obs code to obtain the V/89 filter mock observation in the observation/imaging module of \textsc{cocoa} are listed below: 

\begin{figure}
\begin{center}
 \includegraphics[scale=0.4]{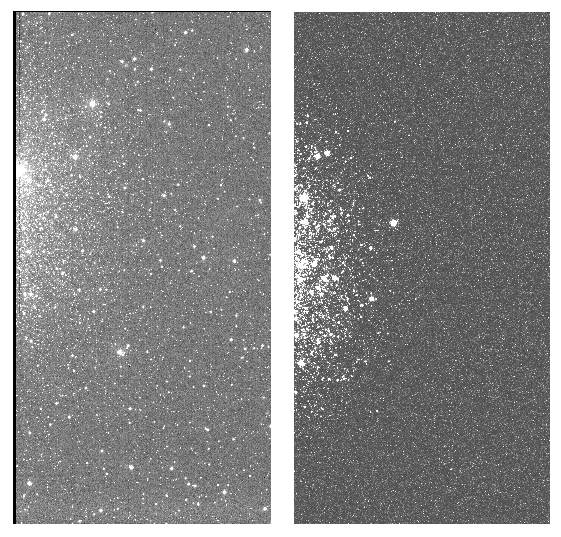} 
 \caption{\textbf{Left:} Observational image of the Galactic GC NGC 2808. The observation was carried out in the V/89 filter of the ESO Wide Field Imager (WFI) that is mounted on the 2.2 meter telescope in La Silla, Chile \citep{ema2011}. \textbf{Right:} Synthetic observational image created using COCOA of a GC model with observational properties compareable to NGC 2808. The parameters for creating this mock observational image were set to replicate the properties of the WFI instrument mounted on a 2.2 meter telescope.}
   \label{ngc2808-cocoa}
\end{center}
\end{figure}

\begin{itemize}
  \item DISTANCE = 9600 [parsecs]
  \item NAXIS1     = 2048 [image width in pixels]
  \item NAXIS2     = 4096 [image length in pixels]
  \item PIXSCALE = 0.238 [arcseconds/pixel]
  \item GAIN = 2.0 [photons/ADU]
  \item SATLEVEL = 100000.0 [counts]
  \item EXPOSURE = 120.0 [multiple direct counts]
  \item SEEING = 0.78 [arcseconds]
  \item BACKGROUND = 1
  \item PSF = M [Moffat distribution for the PSF Model]
  \item M\_BETA = 1.6  [beta parameter for Moffat distribution]
  \item NOISE = 1 [Poisson noise was enabled]
\end{itemize}

For the B/99 band observations, most of the above parameters were the same but the exposure value was changed to 180.0 to match the exposure time of the real B/99 filter observation. After the mock observational images were generated in the two filters, we used the photometry module in \textsc{cocoa} to carry out PSF photometry on the images and obtain catalogues of stars. The recovered instrumental magnitudes were first converted to absolute magnitudes by comparing the instrumental magnitudes with the absolute magnitudes provided by \textsc{GALEVNB}. The absolute magnitudes were then converted to apparent ones using the distance modulus (15.23) and reddening value (E(B-V)=0.17) used by the observers \citep{ema2011}. We then matched the stars from the two different filters based on the recovered position of the stars to obtain a CMD that we could then compare with the observational CMD.

\begin{figure}
\begin{center}
 \includegraphics[width=\columnwidth]{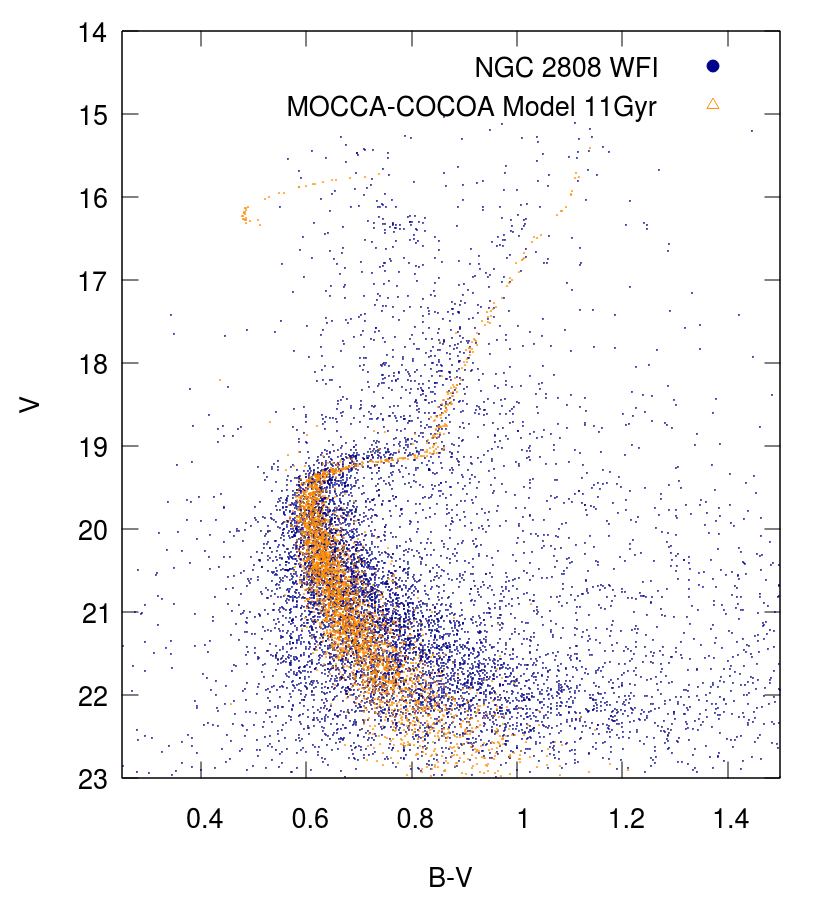} 
 \caption{Colour magnitude diagram showing the comparison between the results of the actual observation of NGC 2808 (blue points) with that of a mock observation of a simulated cluster model at 11 Gyrs with the same metallicity as NGC 2808 (orange points).}
   \label{comp-cmd-ngc2808}
\end{center}
\end{figure}

Figure \ref{comp-cmd-ngc2808} shows the result of this comparison between the observation of NGC 2808 and of a cluster model with the same age and metallicity as NGC 2808. From the comparison it can be seen that the photometry from the mock observation overlaps quite well with the actual observation. In this particular case, the actual observation is plagued by background stars, however the positions of the main sequence (MS) turn-off match quite well in both the CMDs. There are some minor differences with regards to the extension of the subgiant branch in colour and the position of the base of the giant branch. We carefully investigated these differences and attribute them to the properties of the stars provided by the \textsc{sse} and \textsc{bse} codes that use analytic fitting formulae based on the stellar models computed by \citet{pols98}. For low metallicity values typical of globular clusters, the subgiant branch for a set of stars evolved with \textsc{sse} is slightly extended in the redder colour and subsequently the base of the red giant branch also shifts to a redder colour compared to their positions in actual observations and also in CMD isochrones based on newer stellar evolution track models such as \textsc{parsec} isochrones \citep{bressan}. These differences in the stellar properties can have an influence on the global and specific properties of star clusters (such as luminosity function, surface brightness profile) that are obtained from simulated models. Moreover, these differences should be taken into account when making detailed comparisons of simulated models and observations for the purpose of constraining initial conditions for particular globular clusters. The population synthesis code \textsc{sevn} developed by \citet{spera15} reads and interpolates a grid of \textsc{parsec} stellar evolution isochrones to determine stellar properties for massive stars. Such an approach which extends to lower masses can be useful for stellar dynamic codes to get stellar properties based on the latest evolution models. This also demonstrates that using \textsc{cocoa} to compare results of simulated models with actual observations can be helpful in determining and understanding the shortcomings of stellar evolution models and codes. 

\subsection{COCOA \& GALEVNB} \label{sec:galevnb}

COCOA can be used together with the publicly available \textsc{GALEVNB} code \footnote{\textsc{GALEVNB} can be obtained from: \url{http://silkroad.bao.ac.cn/repos/galevnb/}} \citep{kotulla09,pang16} to obtain magnitudes in different photometric filters for stars in simulation snapshots. The \textsc{GALEVNB} code can provide magnitudes for stars in snapshots of numerical simulations in a variety of different filters for various wavelengths. Using stellar properties such as mass, temperature, luminosity and metallicity, \textsc{GALEVNB} can generate a spectra that covers far UV to far IR wavelengths with a resolution of 20 \AA  \citep{pang16}. To obtain magnitudes in different filters, \textsc{GALEVNB} convolves the spectra with the filter response functions and applies selected zero-points corrections.

The projected snapshot generated by \textsc{COCOA} can be used with \textsc{GALEVNB} in order to obtain absolute magnitudes for stars in \textsc{MOCCA} snapshots in different filters, this allows us to simulate multiband observations of simulated cluster models. \textsc{GALEVNB} can also be used by \textit{N}-body modellers (working with \textsc{NBODY6/NBODY6++GPU} to obtain absolute magnitudes for their simulation snapshots in multiple filters. Having absolute magnitudes and projected positions of stars, the imaging and photometry modules in \textsc{COCOA} can be used to create mock observations and photometry on these mock images from \textit{N}-body snapshots. Having magnitudes in instrumental filters is particularly useful as it can allow for meaningful comparison with observational studies and surveys that use particular CMDs to determine binary fractions and multiple population sequences in GCs.

\subsection{Multiband Photometry with COCOA \& Initial Binary Parameter (IBP) Constraints} \label{sec:multiband}

In this section, we show results from multi-band photometry of sparse and moderately dense cluster models that were evolved using MOCCA. The snapshot of the clusters after 12 Gyrs of evolution was used to simulate FITS images in at least 5 photometric bands (U,B,V,R,I). The automated PSF photometry pipeline in \textsc{cocoa} was then used to obtain catalogues of stars from these simulated models. 

All cluster models that were used had initially large N ($1 \times 10^{6}$ stars) and metallicity $z=0.001$. Models beginning with the prefix K have a high initial binary fraction of 95\% and initial binary properties were based on Kroupa's initial binary parameters \citep{kroupa95} . The cluster models beginning with the suffix S have a primordial binary fraction of 5\% (S1) and 10\% (S2). Binary parameters for S models had 'standard' distributions (uniform mass ratio (q) distribution, thermal eccentricity distribution and log-normal or uniform in log semi-major axis distribution). Table \ref{models-info} summarizes the initial and 12 Gyr properties of these cluster models that have have also been described in \citet{belloni1}.

\begin{table*}
\centering
\caption{Table describing the initial properties of the models and their 12 Gyr binary fraction.}
\label{models-info}
\begin{tabular}{clccccc}
\hline
\textbf{Model} & \multicolumn{1}{c}{\textbf{Mass ($M_{\odot}$)}} & \textbf{\begin{tabular}[c]{@{}c@{}}Initial Binary\\  Fraction (\%)\end{tabular}} & \textbf{\begin{tabular}[c]{@{}c@{}}Central Density\\ ($M_{\odot} pc^{-3}$)\end{tabular}} & \textbf{\begin{tabular}[c]{@{}c@{}}$r_{t}$\\ (pc)\end{tabular}} & \textbf{\begin{tabular}[c]{@{}c@{}}$r_{h}$\\ (pc)\end{tabular}} & \textbf{\begin{tabular}[c]{@{}c@{}}12 Gyr Binary \\ Fraction (\%)\end{tabular}} \\ \hline
K1             & $8.07 \times 10^{5}$                            & 95                                                                               & $1.9 \times 10^{2}$                                                                      & 115                                                             & 16.9                                                            & 28.3                                                                            \\
K2             & $8.07 \times 10^{5}$                            & 95                                                                               & $7.8 \times 10^{4}$                                                                      & 115                                                             & 2.3                                                             & 8.9                                                                             \\
S1             & $5.92 \times 10^{5}$                            & 05                                                                               & $2.8 \times 10^{3}$                                                                      & 97                                                              & 7.5                                                             & 4.9                                                                             \\
S2             & $9.15 \times 10^{5}$                            & 10                                                                               & $1.3 \times 10^{5}$                                                                      & 125                                                             & 2.1                                                             & 4.8                                                                             \\ \hline
\end{tabular}
\end{table*}

\begin{figure}
\begin{center}
 \includegraphics[scale=0.4]{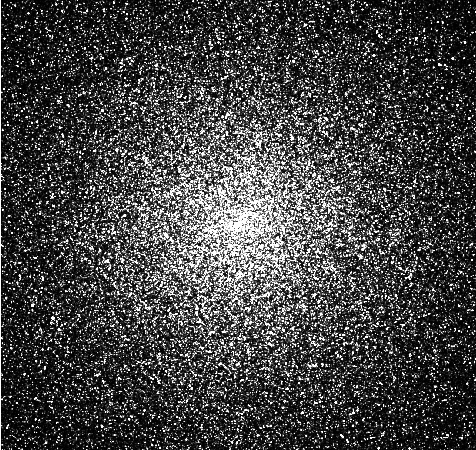} 
 \caption{I-band image of the innermost frame imaging the core of the cluster model K2. 2048 $\times$ 2048 pixels. The cluster was imaged with a high spatial resolution telescope with pixel scale of 0.08 arcsec/pixel. The projected distance to cluster model was 5 kpc.}
   \label{multi-I}
\end{center}
\end{figure}

\begin{figure*}
\begin{center}
 \includegraphics[width=0.48\linewidth]{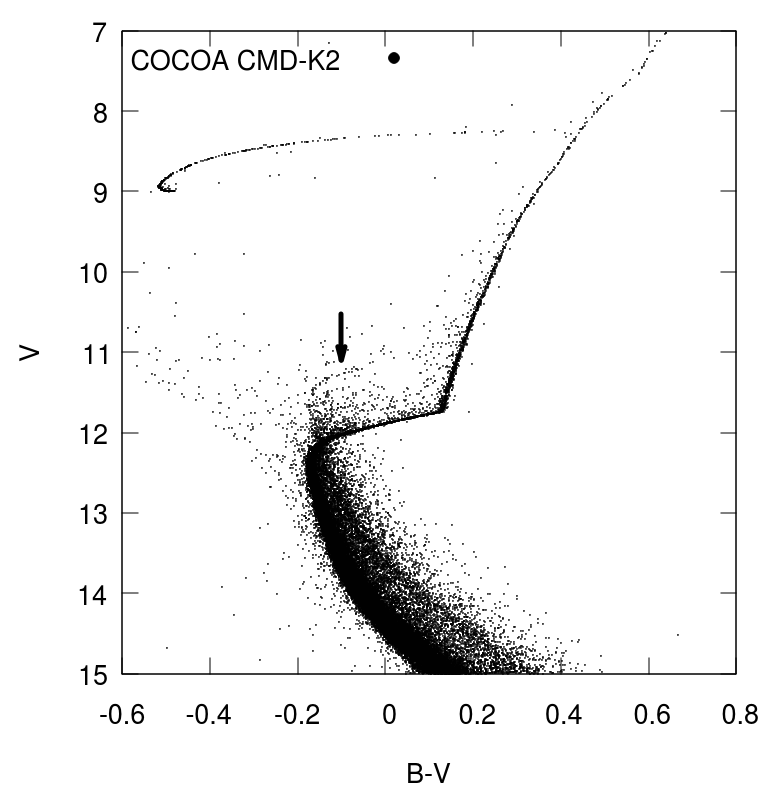}
 \includegraphics[width=0.48\linewidth]{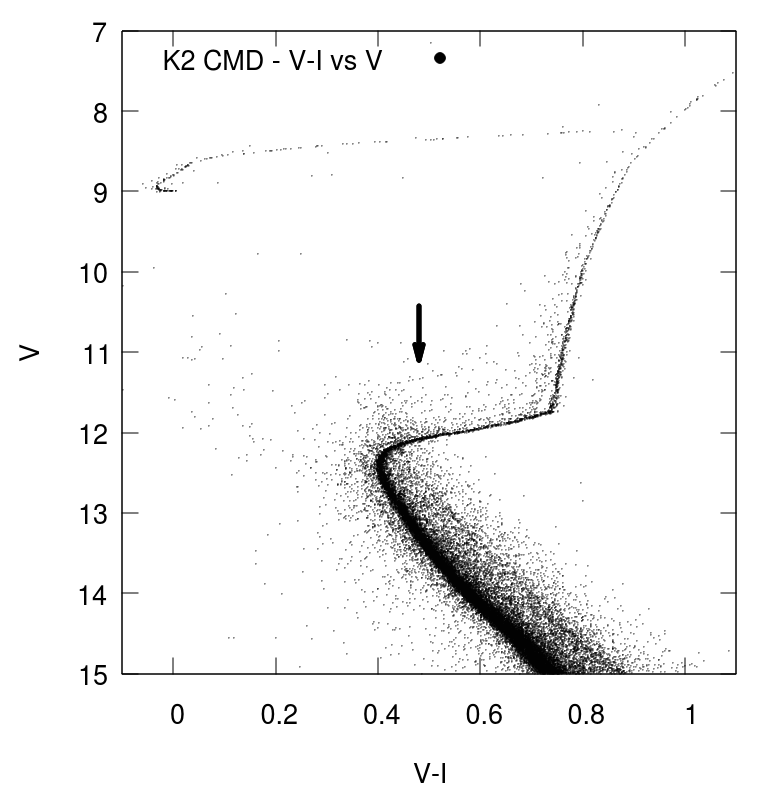}
 \end{center}
 \caption{\textbf{Left Panel:} Shows the B-V vs V CMD for the K2 model using instrumental B and V band magnitudes. There are indications of the presence of a second main sequence due to many binaries with $q$ (mass ratio) close to unity. The black arrows shows these binaries that are visible above the MS turn-off. \textbf{Right Panel:} Shows the V-I vs V CMD for the same model which is significantly denser than the K1 model. Due to this higher density and overall lower binary fraction, the photometric errors are larger and it becomes difficult to distinguish the second sequence. Also with higher density, a significant number of such binaries are disrupted. However, there are still binaries with nearly equal mass components above the MS turn-off that are indicated with the black arrow. }
   \label{cmd-k2}
\end{figure*}

Figure \ref{multi-I} shows the simulated I band image for the moderately dense K2 cluster projected at a distance of 5 kpc and Figure \ref{cmd-k2} shows CMDs for this model. 

\begin{figure*}
\begin{center}
 \includegraphics[width=0.48\linewidth]{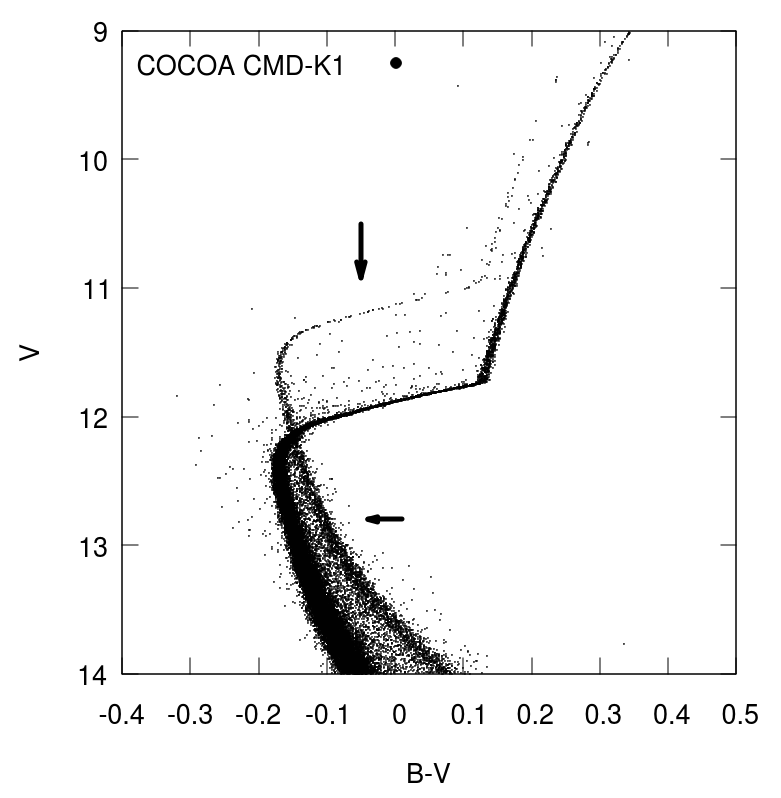}
 \includegraphics[width=0.48\linewidth]{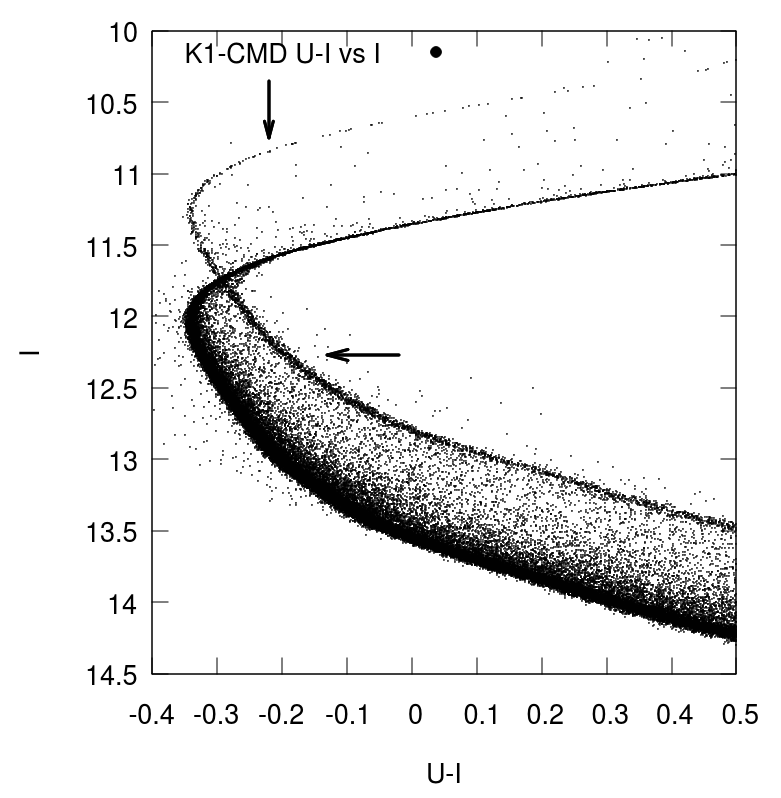}
 \end{center}
 \caption{\textbf{Left Panel:} B-V vs V CMD for the K1 model produced using instrumental B and V band magnitudes. There is a prominent second MS due to many binaries with $q$ close to unity that is indicated with the black arrows.
\textbf{Right Panel:} U-I vs I CMD for the K1 model The figure zooms in on part of the CMD close to the MS turn-off. The presence of a large number of binaries (with $q$ close to 1) causes a second sequence which has its own turnoff 3/4 of a magnitude above the turn-off for single stars.}
   \label{cmd-k1}
\end{figure*}

Multiband simulations were also done for sparse models with high binary fraction and Kroupa distributions for the IBP (K1). As this model was sparse and the initial binary fraction was very high, a lot of binaries survived up to 12 Gyr. The second sequence due to the presence of binaries with $q\approx1$ is very prominent in all CMDs of this model (Figures \ref{cmd-k1}). Binaries with nearly equal mass ratios can result in a second sequence which has an apparent turn off point about 3/4 of a magnitude above the turn off point for single MS stars. The broadening of the MS by binaries of different mass ratios has been thoroughly addressed by \citet{hurleytout98}. As stated above, the second sequence due to the binaries with nearly equal mass components is prominent for models with high primordial binary fractions and initial parameter distributions given by \citet{kroupa1995,marks-kroupa12}. 
This is because this primordial binary distribution of the mass ratio has a disproportionately large number of low mass binaries with $q$ being very close to 1 due to mass feeding during pre-main sequence eigenevolution proposed by \citet{kroupa1995}. This issue with the second-sequence in the initial binary parameters given by \citet{kroupa1995} have been addressed in \citet{belloni17} and a revised prescription for the mass feeding during pre-main sequence evolution has also been proposed.

This second sequence is more prominent when the model is sparse and less prominent in denser models where even primordial binaries with low mass components are more likely to be disrupted. Moreover, the photometric errors are larger in denser fields and the second sequence becomes more difficult to see in the CMD due to these errors. By determining the inferred binary mass ratios from the broadening of the MS, we can determine whether the initial binary parameters we use in the simulated cluster models can recover the observed mass ratio distribution for binaries observed in present day GCs. If the simulated cluster models with particular initial binary fraction and parameter distribution show features which are not corroborated by observational results then these parameters may require further corrections. This is an important application of \textsc{cocoa} which can be helpful in improving theoretical models and initial conditions for star cluster simulations. Apart from second main sequences, other differences between mock observations of star cluster models and actual observations (e.g the number and observational properties of blue stragglers, number of evolved giants) may also be used to improve cluster initial conditions.

The second sequence is also very prominent in the U-I vs I CMD for the K1 model (this can be seen in the right panel of Figure \ref{cmd-k1}). A combination of ultraviolet and visual band filters can be particularly useful for identifying elemental abundances and variations for stellar population in clusters \citep{bellini10}. 

\begin{figure*}
\begin{center}
 \includegraphics[width=0.48\linewidth]{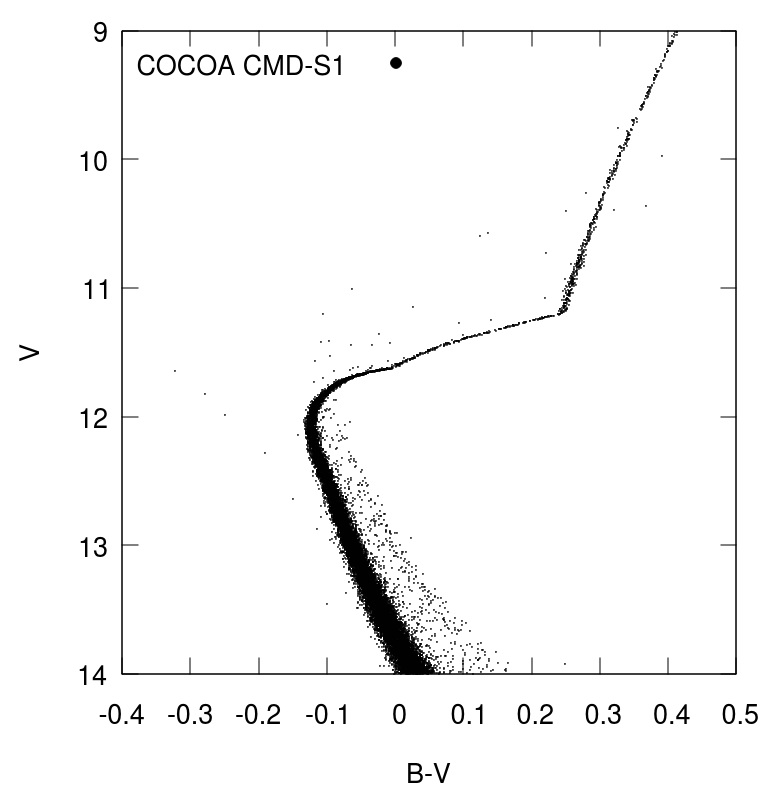}
 \includegraphics[width=0.48\linewidth]{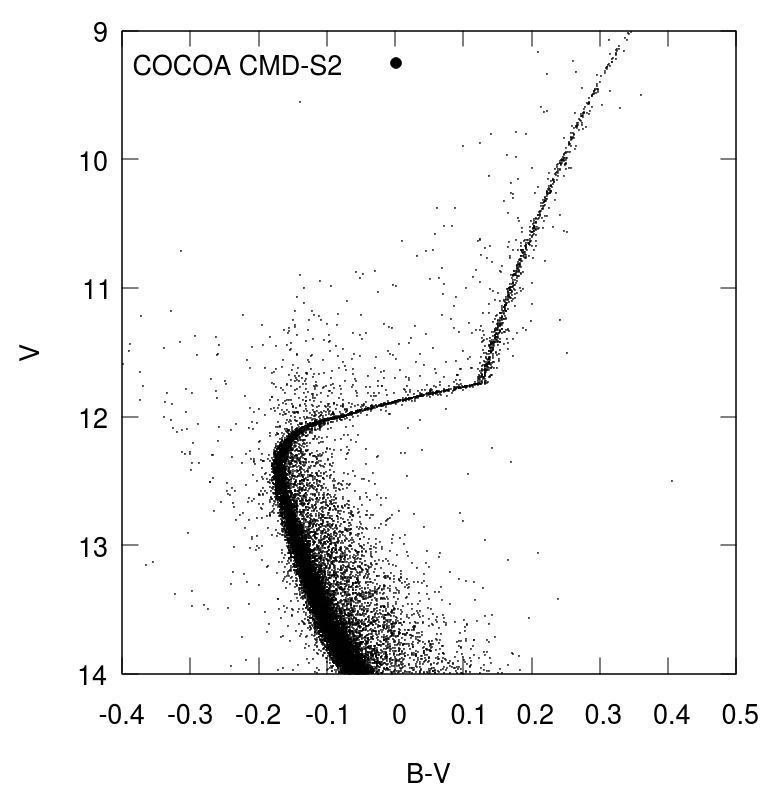}
 \end{center}
 \caption{\textbf{Left Panel:} B-V vs V CMD for the S1 model. Even though this model is initially sparse, no second sequence due to equal mass ratio binaries is seen. CMDs produced using other filters also showed an absence of the second sequence.
\textbf{Right Panel:} B-V vs V CMD for the S2 model. Similar to model S1, there is no presence of a prominent second sequence due to equal mass binaries that was seen in models K1 and K2.}
   \label{s1-s2-cmd}
\end{figure*}

For models with standard binary parameter distribution (S-models), no multiple sequences are seen for CMDs in different filters. B-V vs V CMDs are shown for models S1 and S2 in the Figure \ref{s1-s2-cmd}.

For the sparse model with standard initial binary parameter distributions, The absence of the second sequence due to $q=1$ binaries can be clearly seen in the left panel of Figure \ref{s1-s2-cmd}.

\subsection{Imaging Distant Clusters}

\begin{figure*}
\begin{center}
 \includegraphics[width=0.48\linewidth]{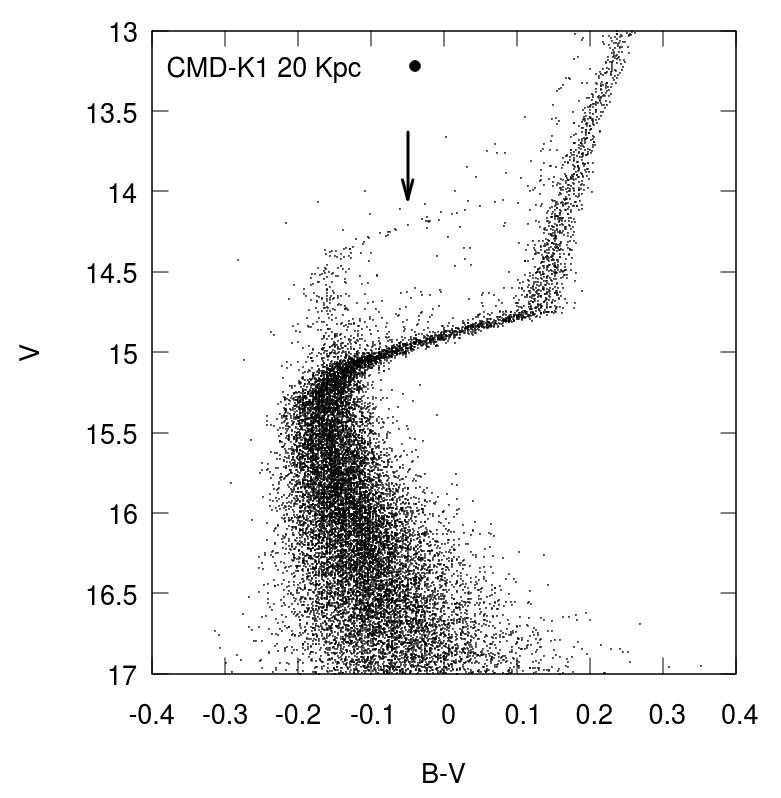}
 \includegraphics[width=0.48\linewidth]{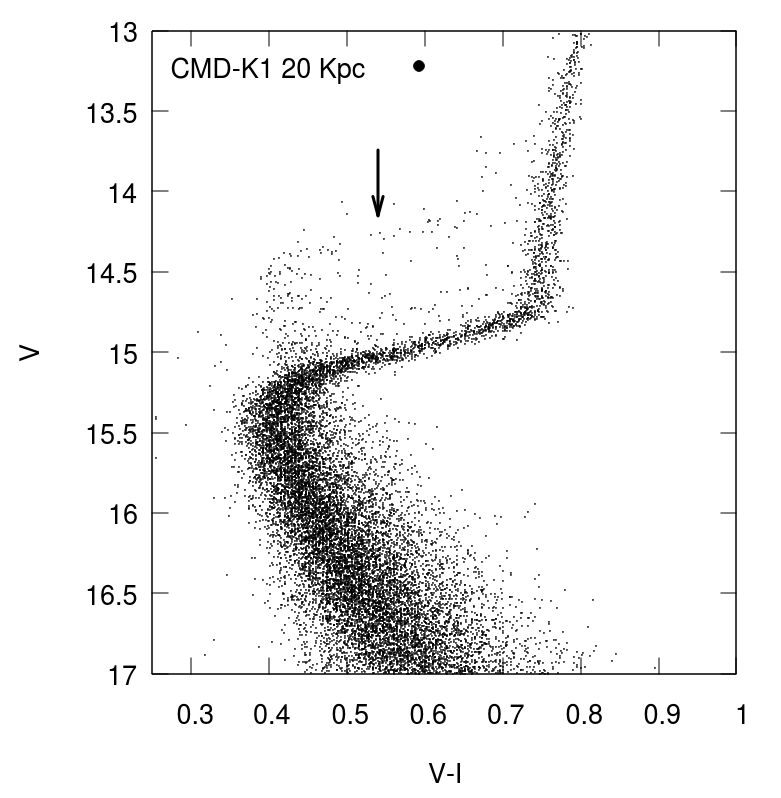}
 \end{center}
 \caption{\textbf{Left Panel:} B-V vs V CMD for the K1 model. The cluster was projected at a distance of 20 kpc and the spatial resolution of the instrument was 0.08 arcsec/pixel. The second sequence due to binaries with nearly equal mass components is visible above the MS turn-off and is indicated with the black arrow.
\textbf{Right Panel:} V-I vs V CMD for the K1 model using instrumental V and I band magnitudes. The second sequence above the MS turn-off is indicated with the black arrow.}
   \label{k1-dist-cmds}
\end{figure*}

In this section, we briefly present results from the K1 cluster model for which mock observations were produced with \textsc{cocoa} at a distance of 20 kpc. Photometry was carried out on these mock observations using \textsc{cocoa} and the CMDs for this model still show a prominent second sequence due to binaries with large mass ratio values. Despite photometric errors being higher when the cluster was observed at a large distance, the second sequence is fairly prominent because the cluster model is quite sparse. Figure \ref{k1-dist-cmds} shows CMDs B-V vs I and V-I vs V CMDs for the K1 model observed at a distance of 20 kpc.

\begin{figure*}
\begin{center}
 \includegraphics[width=0.48\linewidth]{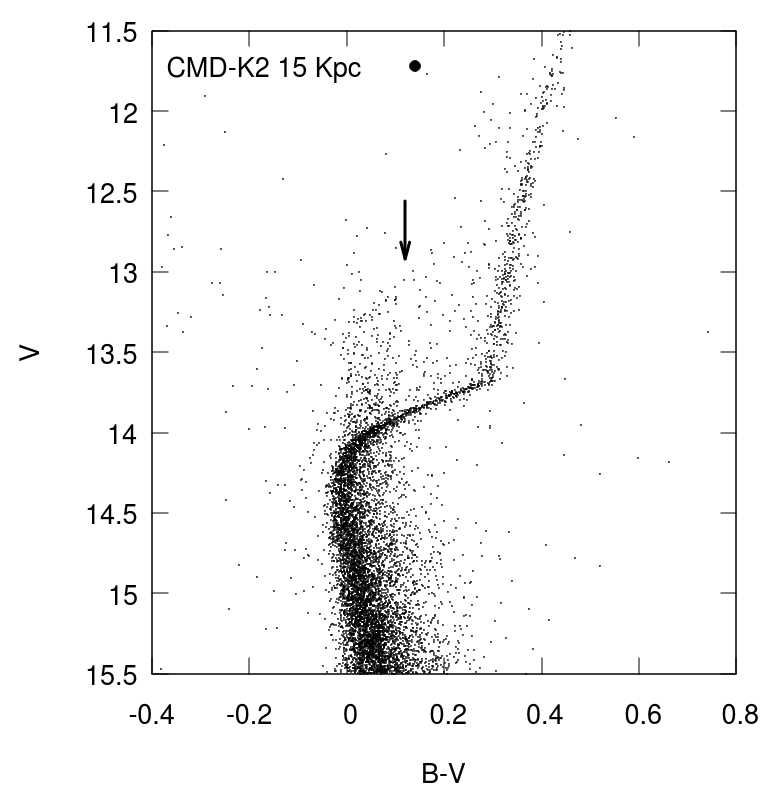}
 \includegraphics[width=0.48\linewidth]{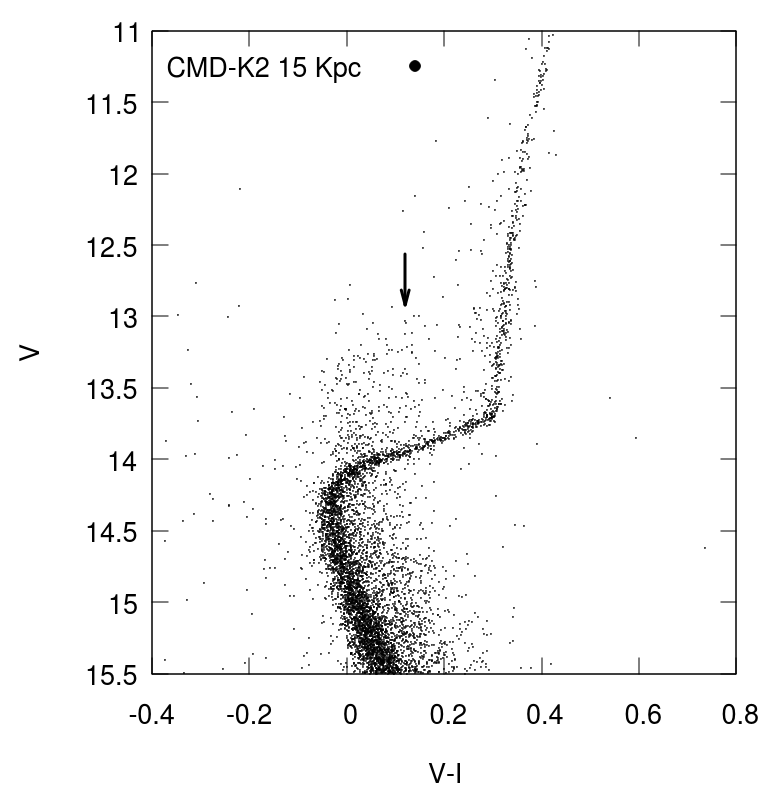}
 \end{center}
 \caption{\textbf{Left Panel:} B-V vs V CMD for the K2 model using instrumental B and V band magnitudes. The cluster was projected at a distance of 15 kpc and the spatial resolution of the instrument was 0.08 arcsecond/pixel. The black arrow indicates the second sequence due to $q\approx1$ binaries above the MS turn-off.
\textbf{Right Panel:} V-I vs V CMD for the K1 model using instrumental V and I band magnitudes. The second sequence above the MS turn-off is indicated with the black arrow.}
   \label{k2-dist-cmds}
\end{figure*}

To check the influence of cluster density and distance on recovering photometric features. We also created mock observations for the more dense K2 model by observing the cluster at a distance of 15 kpc. Photometric errors were higher for this cluster model as the field is more crowded. Due to the density of the cluster and the photometric noise, it can be difficult to determine whether there could be multiple sequences being caused by high mass ratio binaries. The two panels in Figure \ref{k2-dist-cmds} show the B-V vs V and V-I vs V colour magnitude diagrams for the K2 model projected at a distance of 15 kpc. These results show that for dense models, photometric features like a second sequence due to binaries with nearly equal mass components can only be resolved for nearby clusters. For sparse models, such a feature may even be resolved for a distant cluster. This is just to show one of the many possibilities of how the distance to a cluster and its density may influence the observational results. This can be particularly useful for determining whether certain population of stars, binary fractions and other properties can be properly recovered for very distant cluster and how those results would change if that cluster was observed at a closer distance.

\section{Future Development \& Improvements}

There are a variety of uses and applications for the \textsc{cocoa} code and some of these were shown in the previous section. However, there are numerous features that will be further developed to improve comparison between simulations and observations of star clusters. 

The code can already simulate the movement of stars in binary systems on short time scales and generate a sequence of projected snapshots for which observations can be simulated. Accurate treatment of changes in magnitudes for such systems during eclipses will be very useful in estimating the number of eclipsing binaries that maybe observed in particular star clusters. Moreover, this accurate treatment of changes in magnitude due to the movement of the stars in binaries can also be helpful in identifying and observing exotic objects like cataclysmic variables.

For a sequence of projected snapshots generated on short time scales, it will also be possible to use the velocities of all single and binary stars to change their positions according to their proper motions on the plane of the sky. With photometric analysis of these sequence of snapshots, proper motions for the stars in the cluster can be determined. This can be particularly useful for determining global properties of globular clusters. This approach with tracking the movement of the stars in the projected cluster snapshots will also open up the possibility of simulating and detecting occultations of stars by other cluster members.

For future development it will be very important to introduce more automated procedures to as accurately as possible apply observational techniques to our simulated cluster models. For instance, proper aperture correction procedures can be implemented when photometric catalogues from multiple images are being combined. Particularly interesting would be to determine binary fractions in the same way as it is done by observers \citep{milone12}. This will be helpful in determining the accuracy of the observational technique and influences of photometric errors as values obtained from it can be directly compared with simulations. Automated procedure to use observational methods to determine cluster centers from photometric observations can also be applied to our simulated images to determine the validity in determining the cluster center. This is can be particularly important for obtaining accurate velocity dispersion and other profiles. 

It was mentioned in section \ref{sec:compare} that we need to improve the matching criterion for stars from two different filter catalogues. Comparing positions might not be sufficient especially when the field is extremely dense, therefore a more sophisticated routine which can also compare magnitudes will be required to match the stars. It will be useful to have this feature in \textsc{cocoa} to match the stars and produce CMDs automatically for the users.

There are also plans for integrating the \textsc{sisco} \citep{bianchini15} code that can create mock kinematic observations of simulated star cluster models with \textsc{cocoa}. \textsc{sisco} has already been modified to use the projected snapshot from \textsc{cocoa} to produce simulated IFU observations of star cluster models. Both \textsc{cocoa} and \textsc{sisco} were recently used to simulate photometric and kinematic observations of a dark star cluster model in \citet{askar-dark}. Currently, \textsc{cocoa} can generate surface brightness and velocity dispersion profiles (produced from the projected snapshot) that can be fit to simple dynamical models like \citet{king66}. In the future, we plan on introducing options to fit profiles to different models such as Jean's model or more sophisticated multi-mass models such as those developed by \citet{gz2015} \footnote{\textsc{LIMEPY} code developed by \citet{gz2015} can be obtained from: \url{https://github.com/mgieles/limepy}}, \citet{devita} and \citet{peuten17}. With integration of \textsc{sisco}, it is possible to obtain mock observational kinematic profiles which will further improve comparison with observations and can help determine the accuracy of these models in recovering cluster parameters.

\textsc{cocoa} will also be integrated within the Virtual Observatory Star Cluster Analysis service (VOSCA) service \footnote{\url{https://astrogrid-pl.org/vo/user_doc/using_vosca}} which is part of the Astrogrid-PL tools to allow users to create synthetic observations of particular models from a large database of \textsc{mocca} and \textit{N}-body models. Astrogrid-PL \footnote{\url{https://astrogrid-pl.org}} is a Polish project that has been developed to exploit latest computational technology to provides domain-specific grid services and tools that can be used for research in astronomy and astrophysics.

\section{Conclusions}

Creating mock observations of star cluster models and comparing them with observations has numerous advantages. Not only is this useful in seeing how cluster models would appear as observations but has significance in checking the validity of the observer`s data reduction processes. It is also particularly useful for comparing simulated models with real star clusters, which allows one to constrain initial conditions for star cluster models and rule out values that do not match with observations. The purpose of \textsc{cocoa} is to provide theoreticians with an easy to use tool that can help them present their results in a way that is most meaningful for observers. This paper demonstrated the various applications of \textsc{cocoa} and the insights that one can gain from observationally viewing simulated star clusters. Problems with binary parameter distributions, initial conditions and cluster parameters may not be evident from numerical data of evolved clusters. Simply simulating ideal mock observations sheds new light on that data and with theoreticians producing more realistic simulations, the comparison with observations becomes even more necessary. \\textsc{cocoa} can also be significantly important for observers that would like to make use of interesting theoretical results and check whether observations can verify those results. For instance, a number of numerical simulations predict expected number of various exotic objects like cataclysmic variables \citep{belloni1}, blue stragglers \citep{hypki2} and other interesting objects. With \textsc{cocoa}, it is possible to check whether such predicted numbers of exotic objects and their spatial distribution can be observationally recovered. Previous works \citep{sills13,chatter13,pc16,dv17,geller17,sollima17} that have carried out such comparisons between observation and theory have been able to provide important constraints and results. Furthermore, recent \textsc{mocca} simulations by \citet{giersz-imbh} have predicted the formation of an IMBH in GCs, particularly interesting are models in which the IMBH dominates the present-day mass of the cluster. \textsc{cocoa} has been used to check whether such dark star clusters are observed in the Galaxy. In \citet{askar-dark} a comparison for the observed photometric properties of a simulation model was carried out using \textsc{cocoa} which together with mock kinematic observations from \textsc{sisco} \citep{bianchini15} helped to identify NGC 6535 as a possible candidate for harbouring an IMBH.   

Automating observational data reduction and providing end results to users will be particularly useful for theoreticians who are not familiar with observational procedures. With \textsc{cocoa}, all the steps including PSF photometry has been automated. Users with observational experience have the option to use modules that they require and generate mock observational images of the cluster and then perform photometry on them using their preferred tools and scripts. Another motivation for automating the procedures in \textsc{cocoa} is to be able to simulate and analyse mock observations of about two thousand models that were simulated as part of the MOCCA-SURVEY Database project \citep{askar-gw}. More surveys of star clusters models will be carried out with the Monte-Carlo code \textsc{mocca} in the near future. Analysing models with different initial conditions using \textsc{cocoa} can help in improving simulations to reproduce observed cluster models. \textsc{cocoa} will be made publicly available and it can be further developed by the community to incorporate more realistic synthetic observations.

\section*{Acknowledgements}
We are extremely grateful to the reviewer for providing a very thorough report that was very informative and gave suggestions that have been very helpful in improving the presentation, quality and content of the paper.
We would like to thank Ralf Kotulla for making the \textsc{galev} code accessible for generating magnitudes in different filters. We would also like to thank Douglas Heggie for providing the the code that fits the \citet{king66} model to surface brightness and velocity dispersion profiles.
We are very grateful to the following people for providing fruitful discussions, feedback, resolving technical difficulties and explaining observational procedures that helped in developing \textsc{cocoa}: Weronika Narloch, Krystian Ilkiewicz, Arkadiusz Olech, Grzegorz Pietrzy{\'n}ski, Barbara Lanzoni, Ammar Askar, Arkadiusz Hypki, Diogo Belloni and Paolo Bianchini. AA and MG were supported by the National Science
Centre (NCN) through the grant DEC-2012/07/B/ST9/04412. AA would also like to acknowledge support by NCN through the grant UMO-2015/17/N/ST9/02573 and support from Nicolaus Copernicus Astronomical Center's grant for young researchers.








\appendix

\section{COCOA Manual}

In this section we provide a brief manual that explains how to obtain and use the different modules of \textsc{cocoa} and also how option files for photometric procedures are created in the code. 

\subsection{Requirements \& Installation}\label{sec:requirements}

\textsc{cocoa} has been primarily developed in Python 2.7. Subprograms in \textsc{cocoa} have been developed in C and Fortran. The code also requires the GNU scientific library to run the observation module. \textsc{cocoa} can be downloaded from the provided links\footnote{ \url{https://github.com/abs2k12/COCOA} \\or \url{http://users.camk.edu.pl/askar/COCOA-17.tar.gz}}. After, extracting the downloaded file, there is a Makefile that needs to be run in order to compile the necessary subprograms to use \textsc{cocoa} on a UNIX machine. This can be done by using the following command in the terminal:

\begin{lstlisting}[language=bash]
  $ make
\end{lstlisting}

More detailed instructions for using \textsc{cocoa} are provided in the README.md file that comes with the code. We would like to request the users to appropriately acknowledge this paper in case they use \textsc{cocoa} or any of the subprograms and routines provided with it. 

\subsection{Projection Module \& Input File}\label{sec:app-projection}

The projection module in COCOA (see section \ref{sec:projection}) has been primarily designed to work with the snapshot that is output by the \textsc{mocca} code. The module reads each line of the snapshot that among other parameters provides the radial position, radial and tangential velocity of each single star and the center of mass position and velocities for each binary in the system. The module uses this data to obtain the projected position and velocity in x,y and z coordinates. The module also checks whether the object is a single star or a binary, in the case that the object is a binary star, the projected position of the individual stars in the binary with respect to the center of the cluster are resolved. There is also an option to obtain the resolved individual velocities of the binary components with respect to the cluster center or to have the center of mass velocities for those binaries in the projected snapshot that is output by \textsc{cocoa}. In order to obtain the resolved position and velocities of individual stars in binary systems, we use the \textsc{get\_orbit} routine provided in the \textsc{PHOEBE 2.0} code \citep{phoebe1,phoebe2} that primarily deals with physics of eclipsing binaries. The code solves the Kepler equation to determine the position and velocities of binary components at a given time. The description of the output produced by the projection module is provided in the README.md file.

With the projection module it is also possible to generate multiple snapshots that track the movement of the individual binary components in their orbits. The time for these snapshots can be defined by the user in units of days through 2 methods. Either they can provide a final time and the interval at which each snapshot should be generated or through an input file in the which customized values can be provided by the user. In the future, in these multiple snapshots the magnitude of the binary components will be changed during eclipses and subsequent observations can be analyzed to check for variability of those systems and detecting such binaries. The projection module also provides the possibility to generate a rotated projected snapshot. The rotation can be controlled by defining three angles that would define the three dimensional rotation matrix.

The following input parameters need to be set in the input\_projection.py file:

\begin{itemize}
 \item snapshot\_name - Name of the snapshot file that will be projected.
  \item seed - Random number seed that is used to generate angles and distributions that are required for the project. The user can change the seed value to get different projections for the same snapshot.
  \item binary\_velocities - This parameter can take the value of either 1 or 0. If the value is set to 1, then in the output file, the projected individual velocities of stars in binaries are resolved with respect to the center of the cluster. If this value is set to 0 then the the projected center of mass velocities of binary systems are provided in the output file.
  \item intimes - This parameter can take the values 0,1 and 2. When the value is set to 0, only one projected snapshot is generated. If the value is set to 1, then the user can generate multiple snapshots over short observational time that needs to be provided in days. With intimes set to 1, the user provides the end time and the time interval at which each snapshot should be generated. If intimes is set to 2 then the user can provide customized times after which each simulation is generated. To do this, the user needs to provide a data file in the working directory (time.dat) in which each row provides the time in days after which a snapshot will be generated. 
 \item finaltimes - Parameter is only relevant when intimes is set to 1. The value needs to be set for the final time in days at which the multiple snapshot should be generated. 
 \item interval - Parameter is only relevant when intimes is set to 1. The value needs to be set for interval in days at which the multiple snapshots should be generated. 
 \item rotation - Input parameter can take the value of either 0 or 1. In the case the value is 0, only the standard projected snapshot is given. If the value is 1, then a rotated projected snapshot is also provided.
 \item alpha - Parameter is only relevant when rotation is set to 1. The value of the angle needs to be input in units of degrees and this will define the rotation about the x-axis.
 \item beta - Similar to the alpha parameter, this parameter defines the rotation angle in degrees about the y-axis.
 \item gamma - Defines the rotation angle in degrees about the z-axis.
\end{itemize}
 
There are additional input parameters in the input\_projection.py file that can be enabled to call external routines that can generate surface brightness and velocity dispersion profiles from the simulation snapshot and fit the \citet{king66} model to those profiles. By default these options are disabled but can be enabled if the user requires them and comments have been added to the input file that explain how to enable these options. It should be noted that some of these external routines make use of functions taken from standard numerical recipes. Also the results from the fitting routines need to be carefully checked as one may encounter problems with the accuracy when fitting particular GC models.   

Once the input\_projection.py file has been correctly configured. The following command can be used to create the projected snapshot (or multiple snapshots depending on the parameters configured in the input file):

\begin{lstlisting}[language=bash]
  $ python projection.py
\end{lstlisting}

The output generated by the code will be the projected snapshot and a file called param\_rtt.py which contains the name of the projected snapshot and the radial extent of the cluster in pc (this file is required by the imaging module).

\subsection{Imaging/Observation Module \& sim2obs Parameter Files}\label{sec:app-params} 

Once the user has a projected snapshot, essentially three columns from it are needed to create FITS images of the synthetic observation. These are two columns for the projected x,y position of the star in the units of parsec and a third column containing the absolute magnitude in the filter/band for which the observations will be simulated. If the user already has a simulation snapshot with the projected position of the stars and their magnitudes then they can skip the projection module and directly use the imaging/observation module by configuring the input\_observation.py and param\_rtt.py files. 

The following input parameters are required when generating FITS images from a projected snapshot of a star cluster model and they can be configured by the user by editing the input\_observation.py file. Below is a list of the input parameters in this file (variable names in the input file are shown in the parenthesis).

\begin{itemize}
  \item NAXIS1 (valx) - Image/CCD size in x (units of pixel, typical values ~ 2048, 4096).
  \item NAXIS2 (valy) - Image/CCD size in y (units of pixel,typical values ~ 2048, 4096).
  \item DB-NX  (posx) - x-coordinate column in projected snapshot file (value of x position should be in pc).
  \item DB-NY  (posy) - y-coordinate column in projected snapshot file (value of y position should be in pc).
  \item DB-NMAG (dmag) - Magnitude column in projected snapshot file (absolute magnitude).
  \item OBJECT (obje) - Object name for which the mock observation is being generated (any string value may be entered).
  \item DISTANCE (dist) - Distance to the cluster in units of parsecs.
  \item FILTER (filt) - Filter name for which the simulation is being generated. Any string can be entered.
  \item PIXSCALE (pixs)- Gives the pixel scale of the instrument in units of arcseconds/pixel. This defines the spatial resolution of the telescope.
  \item GAIN (gain) - Instrument gain in units of photons/ADU.
  \item SATLEVEL (satl) - Saturation level of the detector in counts.
  \item EXPOSURE (expo) - Exposure for the observation as a multiple of direct counts.
  \item SEEING (seei) - Seeing value for the observation in units of arcseconds
  \item BACKGROUND (back)- Background level of the mock observation. Typical values between 1 and 10.
  \item PSF (psff) - Point spread function (PSF) model for the synthetic observation. Either string values 'G' or 'M' can be entered for this parameter.  G for Gaussian profile and M for Moffat profile. There are no variations in the PSF produced by sim2obs.
  \item M\_BETA (mbeta) - In the case a Moffat PSF is used, this defines the beta parameter for the Moffat distribution.
  \item NOISE (noise) - If set to 1 then this parameter introduces a Poisson noise to the observed FITS image. This parameter can take either the value 0 (no noise) or 1 (Poisson noise).
  \item RA\_OFFSET (raof) - Defines the offset value from the center of the cluster in the x direction. This parameter is useful in imaging different field of the cluster model. Units for offset are in arcseconds.
  \item DEC\_OFFSET (deco) - Defines the offset value from the center of the cluster in the y direction.  Units for offset are in arcseconds.
  \item OVERLAP\_FACTOR - This parameter defines the overlapping factor between adjoining frames when multiple images need to be made to image the entire cluster model. The value must always be less than 1.0. The default value for this parameter is set to 0.97, this means that there is a 3\% overlap between neighboring frames. 
\end{itemize}

There are two editable parameters in the param\_rtt.py file. These are as follows:
\begin{itemize}
  \item rtt - This defines the radial extent of the cluster till which the mock observations will be generated. If you would only like to image the cluster up to a certain radius then you can enter that value for this parameter in units of parsec.
  \item snapname - This variable should contain the name of the projected snapshot file in which you have the three columns that were specified in the input\_observation.py file.
\end{itemize}

Once the input files have been configured. The user can generate the FITS images by running the following command in the terminal:

\begin{lstlisting}[language=bash]
  $ python observation.py
\end{lstlisting}

The code will calculate the number of images it needs to create to image the cluster depending on the input size of the CCD, pixel scale and the radial extent of the cluster. All the images will be generated automatically. The output will be produced in the sub-directory ``fits-files''. The innermost frame will be named 0.0.fits, the naming convention for the output FITS files are based on the coordinates of the grid of the mosaic. First digit corresponds to the x-position of the image in the grid of mosaics and the second digit corresponds to the y-position of the image in that grid (x.y.fits).

\subsection{Photometry Module}\label{sec:phot-mod}

After all the FITS images required to image the cluster have been created. The user can use the photometry module to automatically carry out PSF photometry on each of the image. For the convenience of the user, the output of this routine will concatenate the photometry results from all individual images in the grid, clean the catalogue for overlapping stars in multiple images and provide positions of the stars with respect to the cluster center in units of arcseconds along with the instrumental magnitude and the error in the magnitude. In order to run the photometry module, you will need in the working directory of \textsc{cocoa}, compiled stand-alone versions of \textsc{daophot} and \textsc{allstar} \citep{stetson87,stetson94}. The module also requires the clean\_overlap executable which is created in the working directory when the make command for \textsc{cocoa} is used. In order to run the photometry module the following commands needs to be used in the terminal:

\begin{lstlisting}[language=bash]
  $ python photometry.py
\end{lstlisting}

The module essentially creates wrappers that will then run \textsc{daophot} and \textsc{allstar} on the synthetic FITS images in order to carry out PSF photometry. The module will also generate the option files needed by these two codes based on the values of the FWHM of the synthetic observations. In \textsc{sim2obs}, the FWHM of the simulated image is defined by the seeing value divided by the pixel scale of the instrument (these parameters are defined in the imaging module, see section \ref{sec:app-params}). The prescriptions through which these option files are generated based on the FWHM value is shown in section \ref{sec:phot-pars-dao}.

\subsection{Photometry Options \& Parameter Files}\label{sec:phot-pars-dao}

 For daophot.opt, the following parameters are written in the file:
 \begin{itemize}
  \item FWHM: taken to be the FWHM value determined by the standalone tool
  \item FIT= 1.2 $\times$ FWHM. This is the part of the psf used to determine the fit to each star.
  \item PSF= 2.7 $\times$ FWHM.  PSF value defines the width of a square box inside which PSF will be determined.
  \item READ=0.1, this is the value for the readout noise that is used in our option files. Our simulated FITS images have no readout noise, the minimum value which is acceptable for \textsc{daophot} is used.
  \item GAIN: The value for gain is taken from the sim2obs parameter file that the user entered when generating the FITS images. 
  \item TH=3.5. The threshold value is in the units of standard deviation of the sky noise and is kept at 3.5 or 5 in order to avoid reading the Poisson noise which is simulated in our images. 
  \item AN=1, this parameter determines which analytical PSF model to generate. AN=1 generates a Gaussian PSF model.
  \item LOWBAD $\sim$ 10 to 15. Values between 10 an 15 are used for this parameter which is expressed in sky-sigma units. High value for this parameter prevents reading into the noise. 
  \item HIBAD: This value is also taken from the sim2obs parameter file. The user needs to enter the saturation level of the CCD when generating FITS image with sim2obs. The value entered for saturation level is used as the HIBAD value in the daophot.opt file.
  \item WATCH=-2.0. This suppresses detailed output to the terminal. 
  \item VAR=0 . The synthetic observations do not have a variable PSF across the image, VAR is set to 0.
 \end{itemize}
In the photo.opt, the apertures are defined. There is an option to define one or multiple apertures. For the results shown later in the document. A single aperture is defined:
 \begin{itemize}
  \item A1=1.5 $\times$ FWHM.
  \item IS=4.0 $\times$ FWHM. IS defines the inner radius of sky annulus.
    \item OS=5.0 $\times$ FWHM. OS defines the outer radius of sky annulus.
 \end{itemize}
 For the allstar.opt file, the following parameters are defined
  \begin{itemize}
  \item fit=1.2 $\times$ FWHM. Same value is used as for daophot.opt.
  \item isky=2.0 $\times$ FWHM. IS defines the inner radius of sky annulus in the allstar.opt file.
 \item osky=5.0 $\times$ FWHM. OS defines the outer radius of sky annulus.
 \item watch=0. This suppresses detailed output to the terminal.
 \item redet=1. This parameter allows \textsc{allstar} to redetermine centroids of the star and to improve the quality of fit.
 \end{itemize}

\bsp	
\label{lastpage}
\end{document}